\documentclass[notitlepage,a4paper,aps,prd,onecolumn,superscriptaddress,nofootinbib,groupedaddress]{revtex4-2}

\usepackage{amsmath}
\usepackage{amsfonts,color}
\usepackage{amsmath}
\usepackage{amsthm}
\usepackage{amsmath}
\usepackage{empheq}
\usepackage{amsfonts,color}
\usepackage{amssymb,float}
\usepackage{physics}
\usepackage{booktabs,multirow}
\usepackage{titlesec}

\titleformat{\paragraph}
{\normalfont\normalsize\bfseries}{\theparagraph}{1em}{}
\titlespacing*{\paragraph}
{0pt}{3.25ex plus 1ex minus .2ex}{1.5ex plus .2ex}

\usepackage{amsmath}
\usepackage[utf8]{inputenc}
\allowdisplaybreaks
\usepackage{color}
\usepackage{hyperref}
\hypersetup{
    colorlinks=true,
    linkcolor=blue,
    filecolor=magenta,      
   citecolor=blue
}

\usepackage{enumitem,accents}

\usepackage{tikz}
\usetikzlibrary{shapes.geometric}
\usetikzlibrary{arrows.meta,arrows}

\usepackage[normalem]{ulem}

\newcommand{\lc}[1]{\accentset{\circ}{#1}}

\definecolor{mygreen}{rgb}{0,0.7,0}

\begin{document}
	\title{Spontaneous Scalarization of Black Holes in Gauss-Bonnet Teleparallel Gravity}
	
	\author{Sebastian Bahamonde}
	\email{sbahamondebeltran@gmail.com, bahamonde.s.aa@m.titech.ac.jp}
	\affiliation{Department of Physics, Tokyo Institute of Technology
		1-12-1 Ookayama, Meguro-ku, Tokyo 152-8551, Japan.}
\author{Daniela D. Doneva}
	\email{daniela.doneva@uni-tuebingen.de}
	\affiliation{Theoretical Astrophysics, Eberhard Karls University of T\"ubingen, T\"ubingen 72076, Germany}
	\author{ Ludovic Ducobu}
	\email{ludovic.ducobu@umons.ac.be}
	\affiliation{Department of Mathematics and Computer Science, Transilvania university of Brasov, Brasov, Romania}
	\affiliation{Nuclear and Subnuclear Physics, University of Mons, Mons, Belgium}
	
	\author{Christian Pfeifer}
	\email{christian.pfeifer@zarm.uni-bremen.de}
	\affiliation{ZARM, University of Bremen, 28359 Bremen, Germany}
	
	\author{Stoytcho S. Yazadjiev}
	\email{yazad@phys.uni-sofia.bg}
	\affiliation{Theoretical Astrophysics, Eberhard Karls University of T\"ubingen, T\"ubingen 72076, Germany}
	\affiliation{Department of Theoretical Physics, Faculty of Physics, Sofia University, Sofia 1164, Bulgaria}
	\affiliation{Institute of Mathematics and Informatics, Bulgarian Academy of Sciences, Acad. G. Bonchev St. 8, Sofia 1113, Bulgaria}
	
\begin{abstract}
In this paper, we find new scalarized black holes by coupling a scalar field with the Gauss-Bonnet invariant in Teleparallel gravity. The Teleparallel formulation of this theory uses torsion instead of curvature to describe the gravitational interaction and it turns out that, in this language, the usual Gauss-Bonnet term in four dimensions, decays in two distinct boundary terms, the Teleparallel Gauss-Bonnet invariants. Both can be coupled individually, or in any combination to a scalar field, to obtain a Teleparallel Gauss-Bonnet extension of the Teleparallel equivalent of general relativity. The theory we study contains the familiar Riemannian Einstein-Gauss-Bonnet gravity theory as a particular limit and offers a natural extension, in which scalarization is triggered by torsion and with new interesting phenomenology. We demonstrate numerically the existence of asymptotically flat scalarized black hole solutions and show that, depending on the choice of coupling of the boundary terms, they can have a distinct behaviour compared to the ones known from the usual Einstein-Gauss-Bonnet case. More specifically, non-monotonicity of the metric functions and the scalar field can be present, a feature that was not observed until now for static scalarized black hole solutions.
\end{abstract}
	
\maketitle
	
\section{Introduction}
General Relativity (GR) predicts that in vacuum, there is a unique spherically symmetric solution: the Schwarzschild metric, which is characterised only by the mass $M$ of the gravitational object. Adding electromagnetic fields to the theory, i.e., considering the Einstein-Maxwell case, then the unique spherically symmetric and asymptotically flat black hole solutions are characterized by two parameters, the mass and the charge $Q$ of the gravitational object. Further, by considering a rotating scenario, the angular momentum $J$ appears as an additional parameter in Einstein-Maxwell black hole solutions and measures the amount of rotation of the body. Thus, when one describes stationary and asymptotically flat black holes in GR with minimally coupled electromagnetism, the solutions are uniquely characterised by three parameters ($M,Q,J$) only. This fact is known as Israel’s no-hair theorem~\cite{Misner:1973prb,Chrusciel:2012jk}.

One of the well-studied ways of violating the no-hair theorem is by considering a theory consisting of a scalar field nonminimally coupled to the Gauss-Bonnet invariant~\cite{Kanti:1995vq,Torii:1996yi,Pani:2009wy}, which in the context of scalarized black holes is labeled as scalar Gauss-Bonnet (sGB) gravity. It has been shown that for some coupling functions, there are scalarized black hole solutions where the black hole scalar charge emerges from a spontaneous scalarization process. In this process, the predictions of the theory match the predictions of GR in the weak field regime but differ in the strong field regime. This mechanism was first demonstrated for black holes ~\cite{Doneva:2017bvd,Silva:2017uqg,Antoniou:2017acq} and it shares many similarities with the neutron star scalarization observed in scalar-tensor theories~\cite{Damour:1993hw,Damour:1996ke}. The process for black holes works as follows: The described black hole coincides with Schwarzschild one for weak gravitational fields when the spacetime curvature near the horizon is too weak to source the scalar field. For strong gravitational fields at the horizon, that realizes when the black hole mass $M$ falls below a certain threshold, the Schwarzschild solution becomes unstable, and a non-trivial scalar field emerges. 
Thus, the solution bifurcates to a different black hole solution with scalar hair. This transition is usually smooth in sGB and shares similarities with second order phase transitions. This idea is analog for charged black holes in, e.g., Einstein-Maxwell-scalar gravity ~\cite{Herdeiro:2018wub} or black holes in multi-scalar Gauss-Bonnet theories~\cite{Doneva:2020qww,Staykov:2022uwq}. 
After these findings, several studies have suggested that those scalarized black holes could be important to describe astrophysical scenarios in the strong regime~\cite{Silva:2020omi,East:2020hgw,Kuan:2021lol,Elley:2022ept,Doneva:2022byd,AresteSalo:2022hua}.

All the studies mentioned so far assumed from the beginning that the geometrical description of gravity is based on Riemannian geometry, i.e. that the gravitational field is encoded in a spacetime metric and its canonical torsion free and metric compatible Levi-Civita connection. When one considers a more general geometry as it is done in the metric-affine gravity formalism, one can obtain black hole solutions beyond Schwarzschild (and Kerr in axial symmetry) \cite{Bahamonde:2020fnq,Bahamonde:2022kwg,Bahamonde:2021qjk,Bahamonde:2022meb}. Another alternative and interesting description of gravity can be formulated in terms of tetrads and an independent flat ($R^\lambda{}_{\mu\nu\beta}=0$) metric compatible ($\nabla_{\mu}g_{\alpha\beta}=0$) spin connection which thus possesses only torsion\footnote{Note that, if the curvature $R^\lambda{}_{\mu\nu\beta}$ of the independent connection is vanishing, the curvature $\lc{R}^\lambda{}_{\mu\nu\beta}$ of the Levi-Civita connection associated to the metric by the tetrads is not.}. Those theories are labeled as torsional Teleparallel theories of gravity (or metric Teleparallel theories)~\cite{Aldrovandi:2013wha,Bahamonde:2021gfp}. In that framework, one can formulate an equivalent theory to GR which is known as the Teleparallel equivalent of GR (TEGR) where the action is constructed solely by quadratic contractions of the torsion summed with specific coefficients forming the so-called torsion scalar $T$~\cite{Aldrovandi:2013wha,Maluf:2013gaa}. Since GR and TEGR are dynamically equivalent, their predictions are equivalent and then, one cannot distinguish them by making any classical experiment.

Motivated by the fact that GR might not be the final theory of gravity due to its theoretical and observational challenges (see~\cite{Weinberg:1988cp,Peebles:2002gy,Wong:2019kwg,CANTATA:2021ktz,Nojiri:2017ncd,Addazi:2021xuf,Abdalla:2022yfr} for some of them), one can further follow a similar route as it is done in the Riemannian case and consider modified gravity in the Teleparallel framework. A very interesting aspect of such theories is that they can be interpreted as a gauge theory of translations \cite{Pereira:2019woq}, which is a subcase of Poincar\'e gauge theory \cite{Obukhov:2006gea,Obukhov:2022khx}. The mathematical details of modified Teleparallel theories are still under intense investigation since not all details have been understood from the theoretical point of view yet. The theories retain local Lorentz invariance as symmetry when one transforms simultaneously all the fundamental fields involved, i.e.\ the tetrad and the spin connection~\cite{Krssak:2018ywd,Krssak:2015oua,Golovnev:2017dox,Hohmann:2021dhr}. However, there are suggestions that they also predict strong couplings, i.e., there are some gravitational modes that are strongly coupled in different backgrounds~\cite{Golovnev:2020zpv,Golovnev:2018wbh,Golovnev:2020nln,BeltranJimenez:2020fvy,Blagojevic:2020dyq,Bahamonde:2022ohm}.

One of the simplest and famous ones is  $f(T)$ gravity where one upgrades the TEGR action $T$ to an arbitrary function~\cite{Ferraro:2006jd,Bengochea:2008gz,Ferraro:2008ey}. There have been several applications to that theory in the context of cosmology~\cite{Nunes:2018xbm,Nunes:2016qyp,Cai:2019bdh,Hohmann:2017jao,Wang:2020zfv,Briffa:2020qli} and recently on describing black holes~\cite{Bahamonde:2021srr,Jusufi:2022loj,Bahamonde:2022jue,ElBourakadi:2022anr,Pfeifer:2021njm,Bahamonde:2020snl,Bahamonde:2019zea,Awad:2022fhx}. Other alternative Teleparallel theories have been proposed such as $f(T,B)$ gravity where $B$ is the boundary term relating the Ricci scalar $\mathring{R}$ of the Levi-Civita connection with the torsion scalar $T$~\cite{Bahamonde:2015zma,Escamilla-Rivera:2019ulu,Paliathanasis:2017flf}, or New General Relativity (NGR) \cite{Hayashi:1979qx,Hayashi:1981qx}, where the specific coupling coefficients in the torsion scalar are left arbitrary \cite{Hohmann:2018jso,Hohmann:2022wrk}. In $f(T,B)$ gravity, $f(\mathring{R})$ is obtained as a particular limit, but still, when one considers the purely Teleparallel framework, it might suffer from strongly coupled modes in cosmology~\cite{Bahamonde:2020lsm}. 

Another route to generalize TEGR is to couple one scalar field $\Psi$ minimally or nonminimally to torsion. Since Teleparallel gravity comes with a canonical boundary term $B$.  The simplest way to do this is by taking a Lagrangian like $L=F(\psi)T+G(\psi)B+\partial_\mu \psi \partial^\mu \psi+V(\psi)$ from where $f(T,B)$ gravity can be obtained in a particular scalar-tensor representation, and also other more general Teleparallel theories can be obtained~\cite{Geng:2011aj,Bahamonde:2015hza,Hohmann:2018vle,Hohmann:2018dqh,Hohmann:2018ijr,Zubair:2016uhx}. Furthermore, that theory contains the standard Riemannian scalar-tensor $L=F(\psi)\lc{R}+\partial_\mu \psi \partial^\mu \psi+V(\psi)$ theory by choosing $G(\psi)=-F(\psi)$. 

Recently, there has been found that Teleparallel scalar-torsion theories of gravity, for some coupling functions, have scalarized black hole solutions such as the Bocharova–Bronnikov–Melnikov–Bekenstein solution and other new solutions that do not exist in the Riemannian case~\cite{Bahamonde:2022lvh}. However, all those solutions have a scalar hair not being independent of the mass of the black hole, which is similar to the behaviour of scalarized black holes in the Riemannian theory with non-minimal coupling to the Ricci scalar~\cite{Bekenstein:1995un,Herdeiro:2015waa}. Further, no spontaneous scalarization mechanism has been found for that theory. Therefore, the question of whether Teleparallel gravity could offer a realistic description of scalarized black holes is still open. 

In this paper, our main objective is to obtain the first Teleparallel scalarized black hole solutions with spontaneous scalarization.  To do this, we will formulate a Teleparallel sGB (TsGB) theory constructed from nonminimal couplings between the torsional Gauss-Bonnet invariants and a scalar field. Similarly to what happens in the Riemannian case, those invariants are boundary terms in four dimensions~\cite{Kofinas:2014owa,Bahamonde:2016kba} and therefore, their dynamics would also be nontrivial only when the scalar field is nonminimally coupled to them~\cite{Bahamonde:2020vfj}. As we will show later, our theory can be recast as a combination of the sGB theory plus some additional Teleparallel terms. We will then show that those new contributions can generate spontaneous scalarized black hole solutions triggered by torsion. These could open a new way to study astrophysical objects from non-Riemannian geometries.

This paper is organised as follows. In Sec.~\ref{sec:1} we briefly introduce torsional Teleparallel gravity and the TEGR description of GR whose equations are equivalent to GR. Then, we also discuss the Teleparallel Gauss-Bonnet invariants and how they are related to the Riemannian framework. In Sec.~\ref{sec:3} we present the Teleparallel sGB theory studied in this paper and present its field equations and also its relationship with respect to the sGB theory. Sec.~\ref{sec:4} is devoted to explaining how to work in spherical symmetries for Teleparallel gravity and the corresponding field equations for spherical symmetry for our theory. Sec.~\ref{sec:5} contains the most important part of our manuscript where we showed the existence of scalarized black hole solutions. First, in Sec.~\ref{sec:perturbed}, we found analytical perturbed solutions around Schwarzschild where one notices a scalar charge appearing as an extra parameter. Secondly, in Sec.~\ref{sec:BC}, we studied the behaviour of the field equations at infinity and near the horizons where we found some conditions that we will use for the numerical study. The spontaneous scalarization process for this theory is presented in~\ref{sec:spo} and then in~\ref{sec:num} we numerically solve the equations and find new and novel scalarized black hole solutions. We conclude our main results in Sec.~\ref{sec:concl}. 

In this paper, we work with the metric signature $(+---)$ and we use Latin indices for tangent spacetime and Greek indices for spacetime indices. Further, an upper $\circ$ symbol will be used to denote quantities computed with the Levi-Civita connection (Riemannian case).

\section{Introduction to Teleparallel gravity}\label{sec:1}
The fundamental variables of metric Teleparallel gravity are and a flat \& metric-compatible connection with torsion, with connection coefficients $\omega^a{}_{b\mu}$, and tetrad fields $e^a=e^a{}_\mu \dd x^\mu$ (i.e. orthonormal frames), see; \cite{Aldrovandi:2013wha,Bahamonde:2021gfp}. Gravity is encoded in the torsion tensor defined as the antisymmetric part of the connection
\begin{align}
	T^a{}_{\mu\nu} = 2 \left(\partial_{[\mu}e^a{}_{\nu]} + \omega^a{}_{b[\mu} e^b{}_{\nu]}\right)\,.
\end{align}
Because the connection is both flat and metric-compatible, the spin connection coefficients $\omega^a{}_{b\mu}$ are purely gauge degrees of freedom, and one can always choose a gauge such that this quantity vanishes (Weitzenb\"ock gauge) \cite{Krssak:2018ywd,Krssak:2015oua,Golovnev:2017dox,Hohmann:2021dhr}. This means that it is sufficient to only consider the tetrads as the dynamical variables of the theory (keeping in mind this gauge choice). The metric and its inverse can be reproduced from the tetrads as
\begin{eqnarray}
	g_{\mu\nu} = \eta _{ab} e^a{}_\mu e^b{}_{\nu}  \quad \text{and} \quad g^{\mu\nu}= \eta^{ab}e_a{}^\mu e_b{}^\nu\,,
\end{eqnarray}
while they satisfy  $e^a{}_\mu e_b{}^\mu = \delta ^a_b$, and $e^a{}_{\mu}e_a{}^\nu = \delta _\mu ^\nu$ and $\eta_{ab}=\textrm{diag}(1,-1,-1,-1)$ is the Minkowski metric. 

It is convenient to define the contortion tensor as
\begin{eqnarray}
	K^{\rho}{}_{\mu\nu} & =&\Gamma^{\rho}{}_{\mu\nu}-\lc{\Gamma}^{\rho}{}_{\mu\nu}=\frac{1}{2}\left(T_{\mu}{}^{\rho}{}_{\nu}+T_{\nu}{}^{\rho}{}_{\mu}-T^{\rho}{}_{\mu\nu}\right)\,,
\end{eqnarray}
which measures the difference between the Levi-Civita connection $\lc{\Gamma}^{\rho}{}_{\mu\nu}$ and the Teleparallel connection $\Gamma^{\rho}{}_{\mu\nu}$. Up to quadratic contraction of the torsion tensor, one can build three possible non-parity violating invariants:
\begin{eqnarray}
	T_{\alpha\beta\gamma}T^{\alpha\beta\gamma}\,,\quad T_{\alpha\beta\gamma}T^{\beta\alpha\gamma}\,,\quad T^{\lambda}{}_{\lambda\mu}T_{\beta}{}^{\beta\mu}\,,
\end{eqnarray}
from where one can construct the simplest Teleparallel theories of gravity by considering a Lagrangian formed by a linear combination with arbitrary coefficients $L=a_1 T_{\alpha\beta\gamma}T^{\alpha\beta\gamma}+a_2 T_{\alpha\beta\gamma}T^{\beta\alpha\gamma}+a_3 T^{\lambda}{}_{\lambda\mu}T_{\beta}{}^{\beta\mu}$ \footnote{Which is the Lagrangian defining NGR \cite{Hayashi:1979qx}.}. It turns out that if $(a_1,a_2,a_3)=(\tfrac{1}{4},\tfrac{1}{2},-1)$, one can have a special scalar known as the torsion scalar
\begin{eqnarray}
	T=\frac{1}{4} T_{\alpha\beta\gamma}T^{\alpha\beta\gamma}+\frac{1}{2}T_{\alpha\beta\gamma}T^{\beta\alpha\gamma}-T^{\lambda}{}_{\lambda\mu}T_{\beta}{}^{\beta\mu}\,,
\end{eqnarray}
which differs by a boundary term $B$ with respect to the Levi-Civita Ricci scalar, namely,
\begin{eqnarray}
	\lc{R}=-T+ B = - T + \frac{2}{e}\partial_\mu(e\, T_{\beta}{}^{\beta\mu})\,,\label{RTB}
\end{eqnarray}
where $e=\sqrt{-g}$ is the determinant of the tetrad. Due to the above equation, one can formulate a theory having the same equations as GR by just considering $T$ linearly in an action:
\begin{eqnarray}
	S_{\rm TEGR}=-\frac{1}{2\kappa^2}\int T\,e\,\dd^4x .
\end{eqnarray}
One can then further consider higher order contractions of the torsion tensor. Some interesting scalars that one can consider are the Teleparallel Gauss-Bonnet invariants that, similarly as~\eqref{RTB}, can be related to the Riemannian Gauss-Bonnet invariant $\lc{G}$ as
\begin{eqnarray}
	\lc{G}= T_{G}+B_G\,,\label{GB}
\end{eqnarray}
where the Riemannian part is defined as
\begin{eqnarray}
	\lc{G}=\lc{R}_{\alpha\beta\mu\nu}\lc{R}^{\alpha\beta\mu\nu}-4\lc{R}_{\alpha\beta}\lc{R}^{\alpha\beta}+\lc{R}^2\,,
\end{eqnarray}
and the Teleparallel Gauss-Bonnet invariants are~\cite{Kofinas:2014owa,Bahamonde:2016kba}
\begin{eqnarray}
	T_G&=&\delta^{\mu\nu\sigma\lambda}_{\alpha\beta\gamma\epsilon}K^\alpha{}_{\chi\mu}K^{\chi \beta}{}_\nu K^\gamma{}_{\xi \sigma}K^{\xi \epsilon}{}_\lambda+2\delta^{\mu\nu\sigma\lambda}_{\alpha\beta\gamma\epsilon}K^{\alpha\beta}{}_\mu K^\gamma{}_{\chi\nu}K^{\chi\epsilon}{}_{\xi}K^\xi{}_{\sigma\lambda}+2\delta^{\mu\nu\sigma\lambda}_{\alpha\beta\gamma\epsilon} K^{\alpha\beta}{}_\mu K^{\gamma}{}_{\chi\nu} D_\lambda K^{\chi \epsilon}{}_\sigma\,,\label{TG}\\
	B_G&=&\frac{1}{e}\partial_{\mu}
	\Big[e\delta^{\mu\nu\sigma\lambda}_{\alpha\beta\gamma\epsilon}K^{\alpha\beta}{}_{\nu}\Big(  K^{\gamma}{}_{\xi \sigma}K^{\xi \epsilon}{}_\lambda-\frac{1}{2}\lc{R}^{\gamma\epsilon}{}_{\sigma\lambda}\Big)\Big]\,.\label{BG}
\end{eqnarray}
Here, $D_\lambda K^{\chi \epsilon}{}_\sigma=\partial_\lambda K^{\chi \epsilon}{}_\sigma+(\mathring{\Gamma}+K)^\chi{}_{\beta\lambda}K^{\beta \epsilon}{}_\sigma+(\mathring{\Gamma}+K)^\epsilon{}_{\beta\lambda}K^{ \chi\beta}{}_\sigma-(\mathring{\Gamma}+K)^\beta{}_{\sigma\lambda}K^{\chi\epsilon}{}_\beta$ is the covariant derivative of the general connection. Furthermore, $\delta _{ijkl}^{abcd}=  {\displaystyle {\begin{aligned}4!\delta _{[i}^{a} \delta _{j}^{b}\delta _{k}^{c}\delta _{l]}^{d}\end{aligned}}}$, and then $\delta^{\mu\nu\sigma\lambda}_{\alpha\beta\gamma\epsilon}=e^\mu{}_a e^\nu{}_b e^\sigma{}_c e^\lambda{}_d\delta^{abcd}_{ijkl} e^i{}_\alpha e^j{}_\beta e^k{}_\gamma e^l{}_\epsilon$. It is worth mentioning that the first equation derived for $T_G$ (see Eq. (48) in~\cite{Kofinas:2014owa}) had four pieces, but the authors did not notice that, due to the symmetries of contortion tensor, one of them is always vanishing. Therefore, the scalar $T_G$ only contains three pieces as it is written above. 
It should be noted that in Riemannian geometry (curvature-based theories), the torsion tensor is identically zero but the total curvature tensor $R$ is not. Therefore, the above equation, as it is displayed here, is no longer valid. The above relationship between the Gauss-Bonnet scalars only exists by assuming that the manifold contains torsion and that the general curvature is vanishing.
	
It is well-known that $\lc{G}$ is a topological invariant in four dimensions and it is easy to see that the scalar $T_G$ also has the same property. The extra scalar $B_G$ is always a boundary term (independently of the number of  dimensions). This means that one can add to the TEGR action both $T_G$ and $B_G$, and the dynamics will be just the same as GR. However, if one introduces a scalar field, by allowing some non-minimal couplings between these two terms, they will become dynamical. This property is analogue to the Riemannian Gauss-Bonnet invariant. However, we have shown that in the Teleparallel formulation, the usual Gauss-Bonnet term actually decays into two distinct boundary terms, which can be coupled separately.
	
	\section{Teleparallel Scalar Gauss-Bonnet theory}\label{sec:3}
	There are many studies concerning black holes in the sGB theory, which is constructed by introducing non-minimal couplings between a scalar field and the Gauss-Bonnet invariant $\lc{G}$, namely
	\begin{align}\label{action0}
		\mathcal{S}_{\rm sGB} &=\frac{1}{2\kappa^2}\int \Big[\lc{R}-\frac{1}{2}\beta\, \partial_\mu \psi \partial^\mu \psi
		+\alpha \mathcal{G}(\psi)\lc{G}\Big]
		\, \sqrt{-g}\,\dd^4x
		\,.
	\end{align}
	Depending on $\mathcal{G}(\psi)$, several papers have found that the above theory has scalarized black hole solutions and spontaneous scalarization exists~\cite{Doneva:2017bvd,Silva:2017uqg,Antoniou:2017acq}. 
	
	Following a similar approach for Teleparallel gravity, one can replace $\lc{R}$ for $-T$, see \eqref{RTB}, and introduce non-minimal couplings between the Teleparallel Gauss-Bonnet invariants and the scalar field. By doing this, we have the following Teleparallel sGB theory~\cite{Bahamonde:2020vfj,Bahamonde:2016kba}:
	\begin{align}
		\mathcal{S}_{\rm TsGB} &=\frac{1}{2\kappa^2}\int \Big[-T-\frac{1}{2}\beta\, \partial_\mu \psi \partial^\mu \psi
		+\alpha_1 \mathcal{G}_1(\psi)T_G+\alpha_2 \mathcal{G}_2(\psi)B_G\Big]
		\, e\,\dd^4x
		\,,\label{action}
	\end{align}
	where the coupling functions $\mathcal{G}_1(\psi)$ and $\mathcal{G}_2(\psi)$ depend on the scalar field only. The terms $T_G$ and $B_G$ are boundary terms if they appear linearly in the action. The classical teleparallel boundary term $B$ could also be non-minimally coupled via a term $\alpha_0 \mathcal{G}_0(\psi) B$; however analysing the spontaneous scalarization properties of this term turns out to be more involved (more details on this will be reported in future work), which is why we do not include it in our study here and focus on the Gauss-Bonnet terms.
    If $\mathcal{G}_1$ and $\mathcal{G}_2$ are constants, the theory would be equivalent to the standard Einstein one minimally coupled to a scalar field. From Eq.~\eqref{GB}, one can notice that the above action contains~\eqref{action0} since if  $\mathcal{G}_1=\mathcal{G}_2=\mathcal{G}$ and $\alpha_1=\alpha_2=\alpha$, we recover the last term in~\eqref{action0}, which is the coupling between the Riemannian Gauss-Bonnet and the scalar field $\alpha\mathcal{G}(\psi)\mathring{G}$. In several studies~\cite{Doneva:2017bvd,Silva:2017uqg,Antoniou:2017acq,Doneva:2017duq}, spontaneous scalarization of black holes and compact stars have been found for such theories. Since the above action \eqref{action}, constructed from Teleparallel gravity, is more general than previous theories, as we will show later, the theory we consider has a similar and new scalarization process for black holes.

	\subsection{Discussion on action and field equations}
	The field equations can be obtained by taking variations with respect to the tetrad and the scalar field. Let us remind here that, in Teleparallel gravity, the variation of the action with respect to the flat connection coincides with the antisymmetric part of the tetrad field equations so that one can omit them~\cite{Hohmann:2017duq}.  The tetrad variations, however, are very involved for $B_G$. To simplify this task, one can rewrite the action \eqref{action} using~\eqref{GB} as
	\begin{align}
		\mathcal{S}_{\rm TsGB} 
		&=\frac{1}{2\kappa^2}\int \Big[-T-\frac{1}{2}\beta\, \partial_\mu \psi \partial^\mu \psi
		+\alpha_2 \mathcal{G}_2(\psi)(\mathring{G}-T_G)+\alpha_1 \mathcal{G}_1(\psi)T_G\Big] \, e\,\dd^4x\\
		&=\frac{1}{2\kappa^2}\int \Big[\mathring{R}-\frac{1}{2}\beta\, \partial_\mu \psi \partial^\mu \psi
		+\alpha_2 \mathcal{G}_2(\psi)\mathring{G}+\alpha_3\mathcal{G}_3(\psi) T_G\Big] \, e\,\dd^4x
		\,,\label{action2F}
	\end{align}
	with \begin{equation}
		\alpha_3\mathcal{G}_3(\psi)=\alpha_1 \mathcal{G}_1(\psi) - \alpha_2 \mathcal{G}_2(\psi)\,,
	\end{equation}
	such that one can use the well-known variations of the $\lc{G}$ with respect to the metric. Hereafter, we will use the above reparametrization. There are three important limiting cases appearing from the above action:
	\begin{enumerate}
	    \item $\alpha_3=0$ (or equivalently $\alpha_1 \mathcal{G}_1(\psi)=\alpha_2 \mathcal{G}_2(\psi)$): This theory corresponds to the standard sGB theory.
	    \item $\alpha_2=0$: This theory can be understood as a purely Teleparallel theory, where the dynamics are governed by the 4-dimensional topologically invariant $T_G$ alone. 
	    \item $\alpha_3 \mathcal{G}_3(\psi)=-\alpha_2 \mathcal{G}_2(\psi)$ (or equivalently $\alpha_1=0$): As above, this theory can also be understood as a purely Teleparallel theory and the dynamics is determined by the coupling between the scalar field and the boundary term $B_G$ alone.
	\end{enumerate}
	The second and third cases (which fall under $\alpha_3\neq0)$ are new in the literature and they can only exist when one considers Teleparallel gravity. This means that the scalarization process of those theories does not fall into any of the subcategories studied in previous papers (see the review~\cite{Doneva:2022ewd} where those couplings were not studied nor described). Surely, by the use of \eqref{GB}, one can exchange the teleparallel terms $T_G$ and $B_G$ with each other by the trade off of introducing the usual Levi-Civita curvature Gauss-Bonnet term. However, this is not necessary, and these theories can be described by torsion alone. The other way around, it is not possible to study these theories in terms of curvature alone. Therefore we refer to these theories as ``purely Teleparallel" theories.
    This means that the case $\alpha_3\neq0$ could describe novel scalarized black hole solutions. This is the sector we will focus on in this manuscript. Furthermore, one can also consider theories containing contributions from both $\mathcal{G}_1(\psi)$ and $\mathcal{G}_2(\psi)$, which would have contributions from the Riemannian sector and Teleparallel gravity. 
	
	Let us now derive the field equations for the above model. It is easy to notice that the variation with respect to the metric for the first three terms in the above action are 
	\begin{eqnarray}
		\delta\Big[e\Big(\mathring{R}-\frac{1}{2}\beta\, \partial_\mu \psi \partial^\mu \psi+\alpha_2\mathcal{G}_2(\psi)\delta \lc{G}\Big)\Big]&=& 
		-e\Big[\lc{G}^{\mu\nu}+\frac{1}{4}\beta g^{\mu\nu}\partial_{\rho}\psi\partial^{\rho}\psi-\frac{1}{2}\beta\partial^{\mu}\psi\partial^{\nu}\psi\nonumber\\
		&&+\frac{1}{2e}(\delta_{\rho}^{\mu}\delta_{\lambda}^{\nu}+\delta_{\lambda}^{\mu}\delta_{\rho}^{\nu})
		\eta^{\kappa\lambda\alpha\xi}\epsilon^{\rho\gamma\sigma\tau}
		\lc{R}_{\sigma\tau\alpha\xi}
		\lc{\nabla}_{\gamma}\alpha_2\partial_{\kappa}\mathcal{G}_2(\psi)\Big]\delta g_{\mu\nu}\,,\label{part1}
	\end{eqnarray}
	where $\eta^{\rho\gamma\sigma\tau}
	=\tfrac{1}{e}\epsilon^{\rho\gamma\sigma\tau}$, as usually derived in sGB.
	Since our theory has the tetrads as dynamical variables, we must convert the above equation in terms of variations with respect to the tetrad. This can be easily done by using
	\begin{eqnarray}
		\delta_e g_{\mu\nu} = \eta_{ab}\left( e^b{}_{\mu}\delta e^a{}_{\nu}+ e^b{}_{\nu} \delta e^a{}_{\mu}\right)\,,
	\end{eqnarray}
	yielding
	\begin{eqnarray}
		\delta\Big[e\Big(\mathring{R}-\frac{1}{2}\beta\, \partial_\mu \psi \partial^\mu \psi+\alpha_2\mathcal{G}_2(\psi)\delta \lc{G}\Big)\Big]&=& -e\Big[2\lc{G}^{\mu\nu}+\frac{1}{2}\beta g^{\mu\nu}\partial_{\rho}\psi\partial^{\rho}\psi-\beta\partial^{\mu}\psi\partial^{\nu}\psi\nonumber\\
		&&+\frac{1}{e}\alpha_2(\delta_{\rho}^{\mu}\delta_{\lambda}^{\nu}+\delta_{\lambda}^{\mu}\delta_{\rho}^{\nu})
		\eta^{\kappa\lambda\alpha\xi}\epsilon^{\rho\gamma\sigma\tau}
		\lc{R}_{\sigma\tau\alpha\xi}
		\lc{\nabla}_{\gamma}\partial_{\kappa}\mathcal{G}_2(\psi)\Big]\eta_{ab}e^b{}_\mu \delta e^a{}_\nu\,.\label{part1b}
	\end{eqnarray}
	The variations coming from the extra term in Eq.~\eqref{action2F} that is related to Teleparallel gravity have been found in the context of $f(T_G,B_G)$ gravity. By taking Eq~(59) from~\cite{Bahamonde:2016kba}, one has that the variation of the $T_G$ contribution with respect to the tetrad is\footnote{In this paper, the authors used another convention for the contortion tensor. To transform the notation from~\cite{Bahamonde:2016kba} to our paper, one needs to replace $K_{ijk} \rightarrow K_{jki}$ and $T_G\rightarrow -T_G$}
	\begin{eqnarray}
		\delta(e\mathcal{G}_3T_G)&=&e\mathcal{G}_3 \delta T_{G}+\mathcal{G}_3T_G \delta e\\
		&=& e\Big[-\frac{1}{e}\partial_{\mu}\Big(\eta_{al}(Y^{b[lh]}-Y^{h[lb]}+Y^{l[bh]})e_{h}{}^{\mu}e_{b}{}^{\nu}\Big)-\frac{1}{e}T_{iab}e_{h}{}^{\nu}(Y^{b[ih]}-Y^{h[ib]}+Y^{i[bh]})\nonumber\\
		&&+2\mathcal{G}_3\delta^{mbcd}_{ijkl}e_{d}{}^{\nu}K^{ij} {} _m K^{k} {} _ {eb}\partial_a K^{el} {} _c+\mathcal{G}_3T_G e_a{}^\nu\Big]\delta e^{a}{}_\nu\,,\label{part2}
	\end{eqnarray}
	where we introduced the following tensors:
	\begin{align}
		Y^{b}{}_{ij}& := e \mathcal {G} _ 3 X^b {} _ {ij} - 
		2 \delta^{cabd} _ {elkj}\partial_\mu \Big (e\, \mathcal {G} _ 3 e_d \
		{}^\mu K^{el} {} _c K^k {} _ {ia}\Big)\,, \label{eq:defXY}
	\end{align}
	and
	\begin{eqnarray}
		X^{a}{}_{ij}&=&K_j {}^e {} _b K^k {} _ {fc} K^{fl} {} _d \delta^{abcd} _ {iekl} +  
		K^e {} _ {ib} K^k {} _ {fc} K^{fl} {} _d \delta^{bacd} _ {ejkl} + 
		K^k {} _ {ec} K^{ef} {} _b K_j {}^l {} _d \delta^{cbad} _ {kfil} + 
		K^f {} _ {ed} K^{el} {} _b K^k {} _ {ic}\delta^{dbca} _ {flkj}\nonumber\\
		&&+2 K^k {} _ {eb} K^{el} {} _f K^f {} _ {cd}\delta^{abcd} _ {ijkl} + 
		2 K^{ke} {} _b K_j {}^l {} _f K^f {} _ {cd}\delta^{bacd} _ {keil} + 
		2 K^{el} {} _f K^k {} _ {ib} K^a {} _ {cd}\delta^{fbcd} _ {elkj} + 
		2 K^{fc} {} _d K^
		k {} _ {eb} K^{el} {} _i \eta_ {mj} \delta^{dbma} _ {fckl} \nonumber\\
		&&+2 K^k {} _ {eb}\delta^{abcd} _ {ijkl}\partial_d K^{el} {} _c + 
		2 K^{ke} {} _b \delta^{bacd} _ {keil}\partial_d K_j {}^l {} _c\,.\label{X}
	\end{eqnarray}
	By summing up~\eqref{part1b} and~\eqref{part2}, we find that the field equations for our theory read
	\begin{eqnarray}
		0&=&2\lc{G}_{\beta}^{\nu}+\frac{1}{2}\beta \delta^{\nu}_{\beta}\partial_{\rho}\psi\partial^{\rho}\psi-\beta\partial_{\beta}\psi\partial^{\nu}\psi+\frac{1}{e}\alpha_2(g_{\rho\beta}\delta_{\lambda}^{\nu}+g_{\lambda\beta}\delta_{\rho}^{\nu})
		\eta^{\kappa\lambda\alpha\xi}\epsilon^{\rho\gamma\sigma\tau}
		\lc{R}_{\sigma\tau\alpha\xi}
		\lc{\nabla}_{\gamma}\partial_{\kappa}\mathcal{G}_2(\psi) \nonumber\\
		&&+\alpha_3\Big(\frac{1}{e}\partial_{\mu}\Big[\eta_{al}(Y^{b[lh]}-Y^{h[lb]}+Y^{l[bh]})e_{h}{}^{\mu}e_{b}{}^{\nu}\Big]+\frac{1}{e}T_{iab}e_{h}{}^{\nu}(Y^{b[ih]}-Y^{h[ib]}+Y^{i[bh]})\nonumber\\
		&&-2\mathcal{G}_3(\psi)\delta^{mbcd}_{ijkl}e_{d}{}^{\nu}K^{ij} {} _m K^{k} {} _ {eb}\partial_a K^{el} {} _c-\mathcal{G}_3(\psi)T_G e_a{}^\nu\Big)e^a{}_\beta\,,\label{fieldequations}
	\end{eqnarray}
	where $Y^b{}_{ij}$ is defined by Eq.~\eqref{eq:defXY}.
	
	The modified Klein-Gordon equation for this theory can be obtained by varying the action~(\ref{action2F}) with respect to the scalar field, yielding
	\begin{eqnarray}\label{KG}
		\beta \lc{\square}\psi +\alpha_2 \dot{\mathcal{G}}_2(\psi)\lc{G}+\alpha_3 \dot{\mathcal{G}}_3(\psi)T_G=0\,,
	\end{eqnarray}
	where dots denote differentiation with respect to the scalar field, i.e., $\dot{\mathcal{G}}_i(\psi)=d\mathcal{G}_i/d\psi$. This equation tells us that the Gauss-Bonnet invariants might source the scalar field to induce a scalar hair for the black hole.

	\section{Spherical symmetry}\label{sec:4}
	In this section, we will find the field equations in spherical symmetry. First, we will solve the antisymmetric part of the field equations, and then, the resulting symmetric field equations will be shown.
	
	\subsection{Basic ingredients and antisymmetric field equations}
	Let us now start by assuming spherical symmetry. The most general tetrad satisfying spherical symmetry in the Weitzenbock gauge is~\cite{Hohmann:2019nat}
	\begin{equation}
		e^A{}_{\nu}=\left(
		\begin{array}{cccc}
			C_1 & C_2 & 0 & 0 \\
			C_3 \sin\theta \cos\phi & C_4 \sin\theta \cos\phi & C_5 \cos\theta \cos \phi - C_6 \sin\phi  & -\sin\theta (C_5 \sin\phi + C_6 \cos\theta \cos\phi)  \\
			C_3 \sin\theta \sin\phi & C_4 \sin\theta \sin\phi & C_5 \cos\theta \sin \phi + C_6 \cos\phi  & \sin\theta (C_5 \cos\phi - C_6 \cos\theta \sin\phi) \\
			C_3 \cos\theta & C_4 \cos\theta & - C_5 \sin\theta & C_6 \sin^2\theta\\
		\end{array}
		\right)\label{sphtetrad}\,,
	\end{equation}
	where $C_i=C_i(t,r)$ but hereafter, we will consider the stationary case $C_i=C_i(r)$. This tetrad reproduces the metric
\begin{eqnarray}\label{metric0}
\dd s^2&=&\left(C_1^2-C_3^2\right)\dd t^2 -2 ( C_3 C_4-C_1 C_2)\,\dd t\,\dd r - \left(C_4^2-C_2^2\right)\dd r^2- \left(C_5^2+C_6^2\right)\dd \Omega^2\,,
\end{eqnarray}
	which has an off diagonal term $dr \,dt $. Without losing generality (since spherical symmetry plus stationarity imply staticity), we can choose a coordinate system such that the cross term vanishes. To do this, let us introduce the following reparametrization for the functions
	\begin{eqnarray}\label{eq:ansatz}
		C_1(r)&=&\nu A(r) \cosh\beta(r)\,,\quad C_3(r)=\nu A(r) \sinh\beta(r)\,,\\
		C_4(r)&=&\xi B(r) \cosh\beta(r)\,,\quad  C_2(r)=\xi B(r) \sinh\beta(r)\,,\\
		C_5(r)&=&\chi C(r) \cos\alpha(r)\,,\quad  C_6(r)=\chi C(r) \sin\alpha(r)\,,
	\end{eqnarray}
	with $\{\nu,\xi,\chi\}$ being $\pm 1$. This tetrad gives the metric in the standard form in spherical coordinates:
	\begin{equation}
		ds^2=A(r)^2 \,dt^2-B(r)^2\, dr^2-C(r)^2(d\theta^2+\sin^2\theta d\phi^2)\,.\label{metric}
	\end{equation}
	The first remark to mention is that the first term in $T_G$ (see Eq.~\eqref{TG}) is vanishing in spherical symmetry, meaning that $T_G$ will only contain two non-vanishing parts in spherical symmetry.
	
	Hereafter, we will also assume that the scalar field is static and respects spherical symmetry, meaning that $\varphi=\varphi(r)$.
	The field equations for the theory~\eqref{action} can be split into symmetric and antisymmetric parts (see Eq.~\eqref{fieldequations}). Spherical symmetry given by the tetrad~\eqref{sphtetrad} gives that there are only two non-trivial antisymmetric field equations (see Eq.~\eqref{fieldequations}) which read as
	\begin{eqnarray}
		E_{[\theta \phi]}\propto \alpha_3\dot{\mathcal{G}}_3  \varphi'(\sin\alpha(r))(  \cosh \beta(r)) &=&0\,,\label{anti1}\\
		E_{[tr]}\propto \alpha_3\dot{\mathcal{G}}_3\varphi'(\cos\alpha(r) )( \sinh \beta(r))&=&0\,,\label{anti2}
	\end{eqnarray}
	where primes are differentiation with respect to the radial coordinate $r$.
	If we assume that $\alpha_3\dot{\mathcal{G}}_3  \varphi'\neq0 $ to avoid the trivial cases (Riemannian case or constant scalar field), there are two branches that solve both equations. The first one is when
	\begin{equation}\label{branch1}
		\beta=i\pi \,n_1\,,\quad \alpha= \pi\, n_2\,,\quad n_{1,2}\in\mathbb{Z}\,,
	\end{equation}
	while the second branch is obtained for
	\begin{equation}\label{branch2}
		\beta= \frac{i\pi}{2}+i\pi\, n_3\,,\quad \alpha=\frac{\pi}{2}+\pi\, n_4\,,\quad n_{3,4}\in\mathbb{Z}\,.
	\end{equation}
	Note that the choice of $n_i$ does not affect the tetrad since their choice would only introduce some signs of difference but they can be obtained already  by the different signs introduced in the quantities $\{\nu,\xi,\chi\}=\pm 1$. 
	
	Thus, the first branch (see Eq.~\eqref{branch1}) that solves the antisymmetric field equations has the following real tetrad 
	\begin{align}
		e^{(1)}{}^a{}_\mu =\left(
		\begin{array}{cccc}
			\nu  A & 0 & 0 & 0 \\
			0 & \xi   B \sin\theta \cos \phi & \chi C  \cos \theta   \cos \phi  & -\chi C \sin \theta   \sin \phi  \\
			0 & \xi   B \sin \theta  \sin \phi  & \chi C \cos \theta  \sin \phi & \chi  C \sin \theta  \cos \phi  \\
			0 & \xi   B \cos \theta  & -\chi  C \sin \theta  & 0 \\
		\end{array}
		\right)\label{tetrad1}\,,\quad \{\nu,\xi,\chi\}=\pm 1\,.
	\end{align}
	This tetrad is the same one already found in spherical symmetry in other Teleparallel theories~\cite{Boehmer:2011gw,Ruggiero:2015oka,Bahamonde:2019zea,Bahamonde:2020bbc}. Using this tetrad, the Teleparallel Gauss-Bonnet invariants become
	\begin{eqnarray}
		T_G&=&\frac{8 \left(B^3 A''-3 A' B' C'^2-\xi  B^2 \left(2 \chi   A'' C'+A' \left(\xi  B'+2 \chi   C''\right)\right)+B C' \left(A'' C'+2 A' \left(2 \xi  \chi   B'+C''\right)\right)\right)}{A B^5 C^2}\,,\label{TGa}\\
		B_G&=&-\frac{16 \left(B^2 A''+2 \xi  \chi   A' B' C'-B \left(\xi  \chi   A'' C'+A' \left(B'+\xi  \chi   C''\right)\right)\right)}{A B^4 C^2}\label{BGa}\,,
	\end{eqnarray}
	and they correctly reproduce the standard Gauss-Bonnet invariant
	\begin{eqnarray}
		\lc{G}&=&T_G+B_G=\frac{8 \left(-B^3 A''-3 A' B' C'^2+B^2 A' B'+B C' \left(A'' C'+2 A' C''\right)\right)}{A B^5 C^2}\,.\label{GBstandard}
	\end{eqnarray}
	Notice that the sign of $\nu$ is not important for this first branch, and hence, it can be set to one for simplicity.
	
	The second branch (see Eq.~\eqref{branch2}) has the following complex tetrad:
	\begin{align}
		e^{(2)}{}^a{}_\mu =\left(
		\begin{array}{cccc}
			0 & i \xi   B & 0 & 0 \\
			i \nu  A \sin \theta  \cos \phi  & 0 & -\chi  C \sin \phi & -\chi   C\sin \theta  \cos \theta  \cos \phi  \\
			i \nu  A\sin \theta  \sin \phi & 0 & \chi  C \cos \phi  & -\chi  C  \sin \theta  \cos \theta  \sin \phi  \\
			i \nu  A \cos \theta  & 0 & 0 & \chi  C  \sin ^2\theta  \\
		\end{array}
		\right)\label{tetrad2}\,,\quad \{\nu,\xi,\chi\}=\pm 1\,,
	\end{align}
	while the Teleparallel Gauss-Bonnet scalars for the second tetrad~\eqref{tetrad2} become
	\begin{eqnarray}
		T_G&=&\frac{8 \left(B^3 A''-3 A' B' C'^2-B^2 A' B'+B C' \left(A'' C'+2 A' C''\right)\right)}{A B^5 C^2}\,,\label{TGb}\\
		B_G&=&\frac{16 \left(A' B'-B A''\right)}{A B^3 C^2}\label{BGb}\,,
	\end{eqnarray}
	which also correctly reproduces the Gauss-Bonnet invariant~\eqref{GBstandard}. Notice that the above scalars for the second branch do not depend on either $\{\nu,\xi,\chi\}$ so that the tetrad~\eqref{tetrad2} would have the same dynamics for any sign chosen. This means that, for simplicity, we can just set them to one. 
	
	This complex tetrad also has the same form as the one obtained in~\cite{Bahamonde:2021srr} for $f(T,B)$ gravity and~\cite{Bahamonde:2022lvh} for a Teleparallel scalar-tensor theory constructed from $T$ and $B$ and the scalar field. This means that exactly as in the case of $f(T,B,\psi,X)$ gravity, the Teleparallel sGB theory contains two branches satisfying the antisymmetric field equations and  both tetrads coincide with the corresponding ones in $f(T,B,\psi,X)$ gravity. Thus, we have two sets of symmetric field equations that need to be studied separately. In the majority of the previous studies, it has been found that the complex tetrad has simpler equations and exact black hole solutions~\cite{Bahamonde:2021srr,Bahamonde:2022lvh}so that, in this paper, we will concentrate only on this sector. In principle, one could also study the first branch and follow the same analysis that we will carry during this paper, but we will omit this here since the main objective of our work is to show the existence of scalarized solutions for our theory, and then, it is sufficient to analyse the complex tetrad. We leave the study of the field equations for the real tetrad and search for solutions for future work.
	
	\subsection{Field equations in spherical symmetry for the complex tetrad}\label{sec:complex}
	As mentioned in the previous section, the field equations~\eqref{fieldequations}-\eqref{KG} have two different branches having the same metric. In the current section, we will only show the  spherically symmetric field equations for the complex tetrad since we will only focus on this branch. In this case, there are four equations, but only three of them are independent. The (symmetric) tetrad field equations~\eqref{fieldequations} are given by
	\begin{subequations}\label{fieldeqsym}
		\begin{eqnarray}
			E^t{}_t=0&=&\frac{4B'}{r B^3}-\frac{2}{B^2r^2}+\frac{2}{r^2}-\frac{1}{2B^2} \beta   \psi '^2	-\frac{(\alpha_3\dot{\mathcal{G}}_3+\alpha_2 \dot{\mathcal{G}}_2) \left(8 \left(B^2-3\right) B' \psi'-8 B \left(B^2-1\right) \psi''\right)}{r^2 B^5}\nonumber\\
			&&+\frac{16 \alpha_3\dot{\mathcal{G}}_3 \left(B' \psi'-B \psi''\right)}{r^2 B^3}+\frac{8 \left(B^2-1\right) \psi'^2  (\alpha_3\ddot{\mathcal{G}}_3+\alpha_2 \ddot{\mathcal{G}}_2)}{r^2 B^4}-\frac{16 \psi'^2 \alpha_3\ddot{\mathcal{G}}_3}{r^2 B^2}\,,\label{eq:ttB}\\
			E^r{}_r=0&=&-\frac{4 A'}{r A B^2}-\frac{2}{r^2 B^2}+\frac{2}{r^2}+\frac{\beta }{2 B^2}  \psi '^2+\frac{8 \left(B^2-3\right) A' \psi' (\alpha_3\dot{\mathcal{G}}_3+\alpha_2 \dot{\mathcal{G}}_2)}{r^2 A B^4}-\frac{16 A' \psi' \alpha_3\dot{\mathcal{G}}_3}{r^2 A B^2}
			\label{eq:rrB}\\
			E^\theta{}_\theta=0&=&-\frac{2A''}{A B^2}+\frac{2A' B'}{A B^3}-\frac{2A'}{r A B^2}+\frac{2B'}{r B^3}-\frac{\beta }{2 B^2}\psi '^2-\frac{8 A' \psi'^2 (\alpha_3\ddot{\mathcal{G}}_3+\alpha_2 \ddot{\mathcal{G}}_2)}{r A B^4}\nonumber\\
			&&-\frac{8 (\alpha_3\dot{\mathcal{G}}_3+\alpha_2 \dot{\mathcal{G}}_2) \left(B A'' \psi'+A' \left(B \psi''-3 B' \psi'\right)\right)}{r A B^5}\,,
			\label{eq:thetathetaB}
		\end{eqnarray}
	\end{subequations}
	while the modified Klein-Gordon equation~\eqref{KG} becomes
	\begin{eqnarray}
		E_\psi=0&=&		\beta  \left(\frac{\psi ' \left(r B A'+A \left(2 B-r B'\right)\right)}{r A B^3}+\frac{\psi ''}{B^2}\right)- \frac{(\alpha_3\dot{\mathcal{G}}_3+\alpha_2 \dot{\mathcal{G}}_2) \left(8 \left(B^2-3\right) A' B'-8 B \left(B^2-1\right) A''\right)}{r^2 A B^5} \nonumber\\
		&&+\frac{16 \alpha_3\dot{\mathcal{G}}_3 \left(A' B'-B A''\right)}{r^2 A B^3}\,.\label{eq:psiB}
	\end{eqnarray}
It is interesting to mention that for the case where only $B_G$ appears in the action, which is obtained by setting $\alpha_2\mathcal{G}_2+\alpha_3\mathcal{G}_3=0$, the $E^\theta{}_\theta$ component does not depend on the coupling function, and then the form of the scalar field can be directly obtained from~\eqref{eq:rrB} giving $\beta  r A B\phi'^2=4 \left(r A'+A\right) B'-4 B \left(r A''+A'\right)$. Then, that case has different dynamics than the others.

	\section{Scalarized black holes}\label{sec:5}
	This section will be devoted to studying the spherically symmetric field equations derived in the previous section with the aim to construct asymptotically flat scalarized black hole configurations. To do this, in the first two following sections we will obtain perturbed solutions around Schwarzschild, and then, we will find the asymptotic and near horizon boundary conditions to have black holes. The last two sections will study spontaneous scalarization and find solutions numerically.
\subsection{Perturbed solutions}\label{sec:perturbed}
Following the same idea as~\cite{Julie:2019sab}, it is possible to obtain perturbed solutions to the system that would correspond to a generalization of the standard sGB gravity perturbed solutions. This can be achieved by considering that the contribution for Schwarzschild is small, which effectively is the same as taking $\alpha_i\ll 1$. For doing this, we then take
 \begin{align}
		A(r)^2&=1-\frac{2M}{r}+\epsilon\, a_1(r)+\epsilon^2 \, a_2(r)\,,\quad B(r)^{-2}= 1-\frac{2M}{r}+\epsilon \,b_1(r)+\epsilon^2 \, b_2(r)\,,\label{per1}\\
\alpha_i\mathcal{G}_i(\psi)&=\epsilon\,\alpha_i\,\mathcal{G}_{i}(\psi_\infty)+\frac{\epsilon\,\alpha_i}{M^2}\mathcal{G}'_{i}(\psi_\infty)(\psi-\psi_\infty)+\frac{\epsilon\,\alpha_i}{2M^{4}}\mathcal{G}''_{i}(\psi_\infty)(\psi-\psi_\infty)^2\,,\\
\psi(r)&=\psi_\infty+\epsilon \,\psi_1(r)+\epsilon^2\, \psi_2(r)\,,\label{per2}
\end{align}
where $\epsilon\ll 1$ is a small tracking parameter. 
By taking~\eqref{eq:ttB}-\eqref{eq:psiB} and using the above expansions, we find that the metric functions and scalar field expanded up to second order in $\epsilon$ are given by
\begin{align}
g_{tt}=A(r)^2&=1-\frac{2M}{r}-\frac{\epsilon^2}{\beta}\Big[\frac{C_1}{r}-C_2+\alpha_2^2  \mathcal{G}'_{2}{}^2\tilde{a}_1(r)+\alpha_3{}^2  \mathcal{G}'_{3}{}^2\tilde{a}_2(r)+\alpha_2 \alpha_3  \mathcal{G}'_{2} \mathcal{G}'_{3} \tilde{a}_3(r)\Big]\,,\label{sola}\\
-g_{rr}^{-1}= B(r)^{-2}&=1-\frac{2M}{r}-\frac{\epsilon^2}{\beta}\Big[\frac{C_1-2C_2M}{r}+\alpha_2^2  \mathcal{G}'_{2}{}^2 \tilde{b}_1(r)+\alpha_3^2  \mathcal{G}'_{3}{}^2 \tilde{b}_2(r)+\alpha_2 \alpha_3  \mathcal{G}'_{2} \mathcal{G}'_{3} \tilde{b}_3(r)\Big]\,,\\
		\psi(r)&=\psi_\infty-\frac{\epsilon}{\beta}\Big[\alpha_2\mathcal{G}'_{2} \tilde{\psi}_1(r)+\alpha_3\mathcal{G}'_{3} \tilde{\psi}_2(r)\Big]-\frac{\epsilon^2}{\beta^2}\Big[\alpha_2^2\mathcal{G}'_{2}\mathcal{G}''_{2} \tilde{\psi}_3(r)+\alpha_3^2\mathcal{G}'_{3}\mathcal{G}''_{3} \tilde{\psi}_4(r)+\alpha_3 \alpha_2\mathcal{G}'_{2}\mathcal{G}''_{3} \tilde{\psi}_5(r)\nonumber \\
&\,\,\,\,\,\,\,\,\,\,\,\,\,\,\,\,\,\,\,\,\,\,\,\,\,\,\,\,\,\,\,\,\,\,\,\,\,\,\,\,\,\,\,\,\,\,\,\,\,\,\,\,\,\,\,\,\,\,\,\,\,\,\,\,\,\,\,\,\,\,\,\,\,\,\,\,\,\,\,\,\,\,\,\,\,\,\,\,\,\,\,\,\,\,\,\,\,\,\,\,\,\,\,\,\,\,\,\, +\alpha_3 \alpha_2\mathcal{G}'_{3}\mathcal{G}''_{2} \tilde{\psi}_6(r)\Big]\,,\label{solc}
\end{align}
where $C_i$ are integration constants and the functions $\tilde{a}_i,\tilde{b}_i$, and $\psi_i$ are expressed in the Appendix~\ref{appendix1}. Here, all functions are evaluated at $\psi_\infty$. Note that we have set some integration constants such that the metric does not diverge at the horizon and reabsorbed others in appropriate redefinitions of the time coordinate and the value $\psi_\infty$. This solution represents a scalar-hair-endowed Schwarzschild black hole which is a generalization of the solutions presented in~\cite{Bryant:2021xdh} for the sGB case. This can be seen by taking the case $\alpha_3=0=\psi_\infty,\, \beta=1=M\,,C_1=(49/40) \alpha_2^2 \beta \mathcal{G}'_{2}{}^2,\, C_2=0$ and $\alpha_2\mathcal{G}_2(\psi)=-\alpha f(\psi)$ where the above equations coincide with Eqs.~(2.8)-(2.10) presented in~\cite{Bryant:2021xdh}.  Note also that the integration constants $C_1$ and $C_2$ could also be reabsorbed, respectively, in a redefinition of the mass $M$ and time coordinate in order to make them coincide (up to second order in $\epsilon$) with the ones measured by an asymptotic observer. 

\subsection{Asymptotic and near horizon expansions}\label{sec:BC}
	In this section, we are going to solve the equations at the asymptotical regions, i.e., near the horizon $r_H$ and at $r\rightarrow \infty$. We will follow the same strategy as in~\cite{Antoniou:2017acq}. For the asymptotic expansion, we can take
	\begin{eqnarray}
		(A(r)^2,B(r)^2)=1+\sum_{n=1}^{\infty}\frac{(p_n,q_n)}{r}\,,\quad \psi(r)=\psi_\infty+\sum_{n=1}^{\infty}\frac{d_n}{r}\,,\label{asympt}
	\end{eqnarray}
	and the first coefficients $p_1=-2M$ and $d_1=D$ can be associated with the mass and a scalar charge\footnote{It is worth noting that the term ``charge" is used a bit loosely when designing the parameter $D$. Indeed, just as the mass $M$ of the Schwarzschild black hole or the charge $Q$ of the Reissner-Nordström black hole, the parameter $D$ can be used to classify the static spherically symmetric and asymptotically flat solutions of our system. However, unlike the mass or the electric charge, $D$ is not a conserved charge in the usual sense of the term since there is no conservation law (Gauss-law or Noether current) associated to it. In this respect, a more careful denomination might be ``scalar parameter" but the term ``scalar charge" is widespread in the literature so we should keep using this terminology.}. By expanding the field equations~\eqref{fieldeqsym}-\eqref{eq:psiB} at infinity as~\eqref{asympt} up to fourth order, we find that the metric and scalar field behave as 
	\begin{eqnarray}
		g_{tt}(r)=A(r)^2&=&1-\frac{2 M}{r}+\frac{ \beta  D^2 M+32D\alpha_3\dot{\mathcal{G}}_3(\psi_\infty)}{12 r^3}\nonumber\\
		&&+\frac{D^2\beta  M^2-8 MD (\alpha_3\dot{\mathcal{G}}_3(\psi_\infty)+3\alpha_2 \dot{\mathcal{G}}_2(\psi_\infty))+12D^2 \alpha_3\ddot{\mathcal{G}}_3(\psi_\infty) }{6 r^4}\,,\\
		-g_{rr}(r)=B(r)^2&=&1+\frac{2 M}{r}+\frac{16 M^2-\beta  D^2}{4r^2}+\frac{32 M^3-5\beta  D^2 M-32 D \alpha_3\dot{\mathcal{G}}_3(\psi_\infty)}{4r^3}\nonumber\\
		&&+\frac{1}{r^4}\Big[\frac{\beta ^2 D^4}{16}-8 D^2  \alpha_3\ddot{\mathcal{G}}_3(\psi_\infty)-\frac{13}{3} \beta  D^2 M^2-24 D M \alpha_3\dot{\mathcal{G}}_3+8 \alpha_2 D M \dot{\mathcal{G}}_2(\psi_\infty)+16 M^4\Big]\,,\\
		\psi(r)&=&\psi_\infty+\frac{D}{r}+\frac{D M}{r^2}+\frac{-\frac{\beta ^2}{2} D^3+16 \beta  D M^2-64 M \alpha_3\dot{\mathcal{G}}_3(\psi_\infty)}{12 \beta  r^3}\nonumber\\
		&&-\frac{1}{r^4}\Big[\frac{1}{6} \beta  D^3 M+4\left(\frac{ D^2}{3}+\frac{M^2}{\beta }\right) \alpha_3\dot{\mathcal{G}}_3(\psi_\infty)+\frac{8 D M \alpha_3\ddot{\mathcal{G}}_3(\psi_\infty)}{3 \beta }-2 D M^3-\frac{4 \alpha_2 M^2 \dot{\mathcal{G}}_2(\psi_\infty)}{\beta }\Big]\,.\nonumber \\
	\end{eqnarray}
	Clearly, the above equations generalise the sGB case presented in~\cite{Antoniou:2017acq} (see Eq.~(20)), which is obtained by taking the case $\alpha_3=0$ and  $\alpha_2\dot{\mathcal{G}}_2(\psi_\infty)=-f'(\psi_\infty)$ with $\beta=1$. One notices that the Teleparallel contributions appear at a lower order than in the sGB case.
	
	Let us now take expansions near the horizon $r_H$ as
	\begin{eqnarray}\label{expanhorz}
		A(r)^2&=&a_1(r-r_H)+a_2(r-r_H)^2+\dots\,,\\
		B(r)^{-2}&=&b_1(r-r_H)+b_2(r-r_H)^2+\dots\,,\\
		\psi(r)&=& \psi_H+\psi_H'(r-r_H)+\psi_H''(r-r_H)^2+\dots\,.
	\end{eqnarray}
	By assuming these types of expansions, we ensure the fact that $\textrm{det}(g_{\mu\nu})$ is finite at the horizon as long as $b_1$ is non-vanishing.
 
 One can easily solve the system~\eqref{fieldeqsym}-\eqref{eq:psiB} in this region, which, for the scalar field evaluated at the horizon, gives us two possible branches. The first branch is when $\alpha_3\dot{\mathcal{G}}_3\neq\alpha_2 \dot{\mathcal{G}}_2(\psi_H)$, where the scalar field at the horizon is 
	\begin{eqnarray} \label{eq:BC1}
		\psi'_{H}&=& \frac{r_H }{4 (\alpha_2 \dot{\mathcal{G}}_2-\alpha_3 \dot{\mathcal{G}}_3)}\left(1\pm\frac{1}{\beta}\Big[\beta ^2+\frac{32 (\alpha_3 \dot{\mathcal{G}}_3-\alpha_2 \dot{\mathcal{G}}_2) }{r_H^8}\left\{32 \alpha_3^2 \dot{\mathcal{G}}_3^2 (\alpha_3 \dot{\mathcal{G}}_3-\alpha_2 \dot{\mathcal{G}}_2)+\beta  r_H^4 (3 \alpha_2 \dot{\mathcal{G}}_2+\alpha_3 \dot{\mathcal{G}}_3)\right\} \Big]^{1/2}\right)\nonumber \\
		&&-\frac{8 \alpha_3 \dot{\mathcal{G}}_3}{\beta  r_H^3}\,.\label{rhprime}
	\end{eqnarray}
In the above equations, the functions are evaluated at $r_H$. In this expression we have to choose the minus sign since only in this case we can recover the Schwarzschild solution in the limit of vanishing scalar field $\psi_{H} \rightarrow 0$. Note that when the argument of the square root in the equation above becomes negative, regular black hole solutions with scalar hair can not exist. We will call this, the regularity condition. Note that for the sGB case $\alpha_3=0$, Eq.~\eqref{rhprime} becomes
	\begin{eqnarray}
		\psi'_{H}&=&\frac{1}{4 \alpha_2 r_H\dot{\mathcal{G}}_2}\Big[r_H^2\pm \sqrt{r_H^4-\frac{96 \alpha_2^2 \dot{\mathcal{G}}_2^2}{\beta }}\Big] \,,
	\end{eqnarray}
	which matches the result presented in~\cite{Antoniou:2017acq} (Eq.~(14)) after taking the limit $\alpha_2=-1, \mathcal{G}_2(\psi)=f(\psi)$ and $\beta=1$. 

The second branch appears when $\alpha_3\dot{\mathcal{G}}_3=\alpha_2 \dot{\mathcal{G}}_2(\psi_H)$, where this quantity becomes 
	\begin{eqnarray} \label{eq:BC2}
		\psi'_{H}=
		\displaystyle\frac{8 \alpha_2\dot{\mathcal{G}}_2(\psi_H)}{\beta  r_H^3}\,.
	\end{eqnarray}
	 The boundary condition above, contrary to \eqref{eq:BC1}, can be always satisfied because of the disappearance of the square root. This means that for the second branch, there is not a minimum horizon radius needed to obtain scalarised black hole solutions. This feature does not exist in both the sGB case and the first branch due to the appearance of the square root in~\eqref{rhprime}.
 
Clearly, when solving the field equations numerically, the regularity conditions above will serve as an initial condition for the first derivative of the scalar field at the horizon. Thus, depending on the relative strength of both coupling, one should use either Eq.~\eqref{eq:BC1} or \eqref{eq:BC2}.
	
	This analysis suggests that there are two different branches in Teleparallel sGB having asymptotically flat scalarized black hole configurations. This will be studied further in the next sections by analyzing the numerics of the field equations.

\subsection{Spontaneous scalarization}\label{sec:spo}
Spontaneous scalarization provides a mechanism such that, in the weak field regime, the black hole solution is described by Schwarzschild, and, in the strong regime, the scalar field triggers a hair that would be manifested in the fact that the Schwarzschild solution would become unstable. The simplest way to study the spontaneous scalarization process is by considering deviations from the Schwarzschild metric by choosing the metric to have the form $ds^2=e^{\delta(r)}A(r)dt^2-\tfrac{1}{A(r)} dr^2-r^2d\Omega^2$ with $\delta(r)\ll 1$ and $A(r)=1-2M/r$. Further, the perturbation of the scalar field can be written as $\delta\psi(t,r,\theta,\psi)=\tfrac{u(r)}{r} \, e^{-i \omega t}\,Y_{lm}(\theta,\psi)$, which allows us to decouple the tetrad field equations from the scalar field equations. By plugging those expressions into the scalar field equation~\eqref{KG} and after expanding them up to first order, we obtain the following Schrodinger-like form equation,
	\begin{eqnarray}
		\frac{d^2u}{dr^2_{*}} + [ \omega^2 - U(r)]u=0\,, \quad \textrm{with}\,\, 	\quad \mathcal{G}_i'(\psi_0)=0\,,
	\end{eqnarray}
	where we have introduced tortoise coordinates $dr_{*}=(1-2M/r)^{-1}dr$. The potential $U(r)$ for the complex tetrad~\eqref{tetrad2} behaves as
	\begin{eqnarray}
		U(r)=\Big(1-\frac{2M}{r}\Big)\Big[\frac{2M}{r^3}+\frac{l(l+1)}{r^2}-\frac{32M}{r^5\beta} \alpha_3\ddot{\mathcal{G}}_3(\psi_0)+\frac{48M^2}{r^6\beta}(\alpha_3\ddot{\mathcal{G}}_3(\psi_0)+\alpha_2 \ddot{\mathcal{G}}_2(\psi_0))\Big]\,.
	\end{eqnarray}
Here $\psi_0$ is assumed to be the constant Schwarzschild geometry background value of the scalar field. A sufficient condition for having  an unstable mode is~\cite{Doneva:2017bvd,doi:10.1119/1.17935}
	\begin{eqnarray}
		\int_{-\infty}^{+\infty} U(r_{*}) dr_{*}=\int^{\infty}_{2M} \frac{U(r)}{1-\frac{2M}{r}}dr <0\,.
	\end{eqnarray}  
	Therefore, the theory gives us the possibility to have such an unstable mode if
	\begin{eqnarray}\label{eq:BifurcationCondition}
		\frac{-4 \alpha_3\ddot{\mathcal{G}}_3(\psi_0)+6\alpha_2 \ddot{\mathcal{G}}_2(\psi_0)+5 \beta M^2}{20 \beta  M^3}<0\,.
	\end{eqnarray}
	If we choose the sGB case ($\alpha_3=0$) and take $\mathcal{G}_2(\psi_0)=f(\psi_0),\, \beta=4,\alpha_2=-1$ with $\ddot{f}(\psi_0)=\lambda^2$, we recover the condition $M^2<3\lambda^2/10$ as in the standard sGB case considered in~\cite{Doneva:2017bvd}. Thus, spontaneous scalarization can occur in our theory for a much larger choice of parameters for different masses. 
	
	Another important point is that, contrary to the sGB case, where scalarization of non-rotating black holes was possible only for $\alpha_2<0$, we can have scalarization for different signs of the coupling parameters. This is evident from Eq.~\eqref{eq:BifurcationCondition} according to which scalarization is determined not only by the signs of $\alpha_2$ and $\alpha_3$ but also by the relative strength between them. This has interesting implications for the behavior of the numerical solutions as we will see below.
	
	\subsection{Numerical results}\label{sec:num}

	Without loss of generality, we can assume $\beta=4$ that coincides with the convention adopted in \cite{Doneva:2017bvd} and set the background value of the scalar field to zero ($\psi_0=0$). Apart from that, the free parameters we are left with are $\alpha_2$ and $\alpha_3$, and the choice of the coupling functions $\mathcal{G}_2$ and $\mathcal{G}_3$. In order to avoid having to deal with too many parameters simultaneously, while still being able to demonstrate the existence and behavior of solutions, we have decided to fix the two coupling functions in the following form:
	\begin{equation}\label{eq:G2_G3_choice}
		\mathcal{G}_2(\psi)=\frac{1}{12}\left( 1-e^{-6 \psi^2}\right) = \mathcal{G}_3(\psi)\,.
	\end{equation}
	This exponential function has one of the desired properties for scalarization, namely, it allows the GR solutions with a zero scalar field to be also solutions of the more general system of equations in Teleparallel gravity \eqref{fieldeqsym}--\eqref{eq:psiB}. The additive constant is chosen for convenience such that $\mathcal{G}(0)=0$ but it does not affect the field equations since there only the derivatives of $\mathcal{G}_2$ and $\mathcal{G}_3$ enter. In addition, its second derivative at $\psi=0$ is equal to one that is a convenient normalization. More importantly, for this coupling, it was proven in \cite{Doneva:2017bvd,Blazquez-Salcedo:2018jnn} that stable scalarized black hole solutions exist in the sGB case. This is an important property not present for some other coupling functions employed in the literature \cite{Silva:2017uqg,Blazquez-Salcedo:2018jnn}. Having fixed $\mathcal{G}_2$ and $\mathcal{G}_3$, the only theory parameters left to vary are $\alpha_2$ and $\alpha_3$ and more precisely, their relative weight. As already commented, when $\alpha_3=0$ the problem coincides with the sGB case considered in \cite{Doneva:2017bvd}. 
    The following two cases are especially interesting since they are purely Teleparallel; i.e.\ \emph{the scalarization is triggered by torsion}:
    \begin{enumerate}
        \item $\alpha_2=0$ while $\alpha_3\neq 0$, the modification of GR comes only from the Teleparallel topological invariant $T_G$;
        \item $\alpha_2 \mathcal{G}_2 + \alpha_3 \mathcal{G}_3 =0$, the modification of GR comes only from the additional Teleparallel boundary $B_G$.
    \end{enumerate}
    In either of these two limiting cases, no contribution of the Riemannian Gauss-Bonnet term is present. As commented, these cases go beyond the classification of theories allowing for scalarization that is discussed in \cite{Doneva:2022ewd}.
 
        \subsubsection{Numerical setup}
    The black hole solutions are obtained after solving the system of equations \eqref{fieldeqsym}--\eqref{eq:psiB} together with the boundary conditions at the horizon and infinity given in Sec. \ref{sec:BC}. The numerical approach follows the methodology in \cite{Doneva:2017bvd} where the equations \eqref{fieldeqsym}--\eqref{eq:psiB} are transformed to a system of two second order equations -- one for the tetrad function $A(r)$ and one for the scalar field $\psi(r)$. The function $B(r)$ on the other hand can be algebraically related to $A(r)$, $\psi(r)$, and their derivatives. It should be noted that the tetrad functions $\{A(r),B(r)\}$ appear squared in the metric; i.e., $\{g_{tt}(r),-g_{rr}(r)\}=\{A(r)^2,B(r)^2\}$. In the following, we will analyse/plot the tetrad functions $\{A(r),B(r)\}$ instead of their metric counterpart since their behaviour is effectively the same. 
    
    The horizon radius is the black hole input parameter that determines the solution (for fixed theory parameters). We use a shooting method to solve the so-constructed boundary value problem with a shooting parameter being the value of $\psi$ at the horizon. This parameter is determined by the requirement for asymptotic flatness at infinity, namely $\psi(r\rightarrow \infty) = 0$. For the numerical integration of the differential equations, we used a fourth-order Runge-Kutta method. 
	
	An important property of the system of equations is the fact that the requirement for regularity of the metric (or tetrad) functions and the scalar field at the horizon results in an additional boundary condition for the first derivative of the scalar field at the horizon. Depending on the relative strength of both coupling functions, two options exist. Namely, for $\alpha_3\dot{\mathcal{G}}_3 \ne \alpha_2 \dot{\mathcal{G}}_2$  the condition is given by \eqref{eq:BC1}, which we call boundary condition one (BC1), while for  $\alpha_3\dot{\mathcal{G}}_3 = \alpha_2 \dot{\mathcal{G}}_2$ we have \eqref{eq:BC2}, called later boundary condition two (BC2). 
	\subsubsection{Branches of scalarized black holes}

		\begin{figure}[h]
		\includegraphics[scale=0.35]{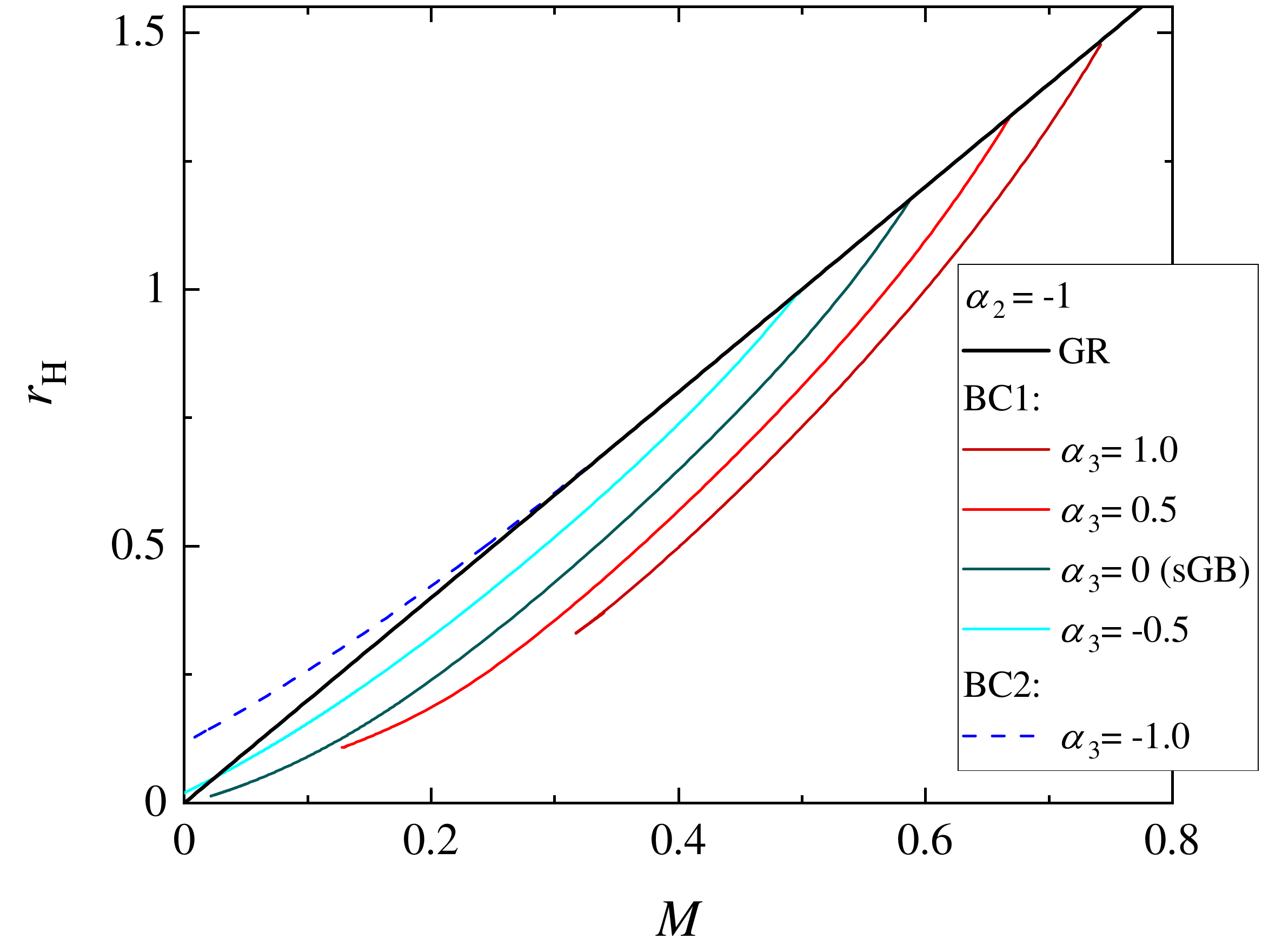}
		\includegraphics[scale=0.35]{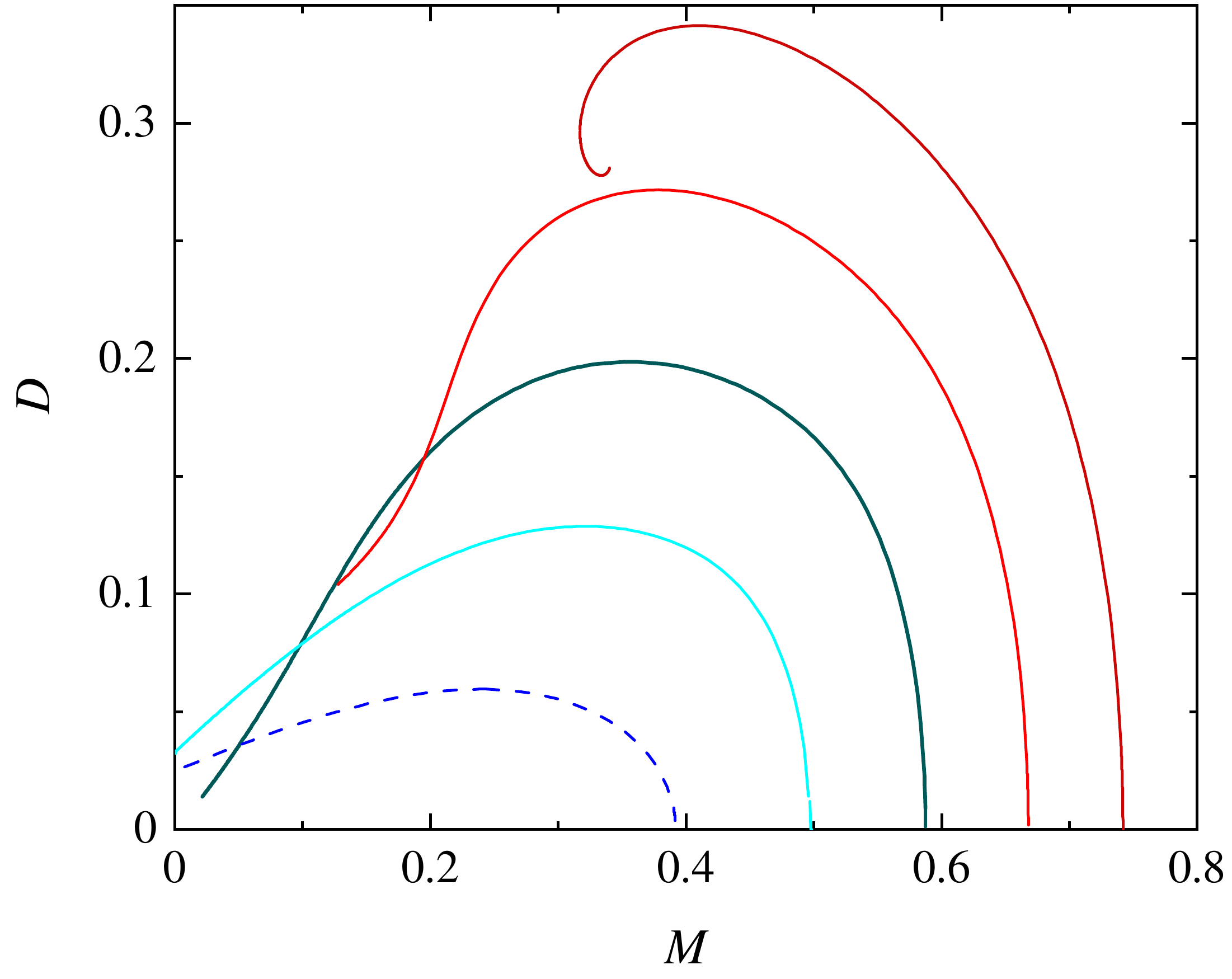}
		\caption{The radius of the horizon (left panel) and the scalar charge (right panel) as functions of mass for a fixed $\alpha_2=-1$ and varying $\alpha_3$. The bald Schwarzschild black hole is depicted with a solid black line, the scalarized branches with $\alpha_3\dot{\mathcal{G}}_3(\psi_H) \ne \alpha_2 \dot{\mathcal{G}}_2(\psi_H)$ and using BC1 with solid lines of different colors, while the solutions with $\alpha_3\dot{\mathcal{G}}_3(\psi_H) = \alpha_2 \dot{\mathcal{G}}_2(\psi_H)$  and BC2 are marked by a dashed line. The pure sGB case (when $\alpha_3=0$) is depicted with a solid dark green line.}
		\label{fig:D_rh_M_var_a3}
	\end{figure}

	 In order to gain some intuition about the existence and behavior of black hole solutions let us start with discussing the bifurcation point, which corresponds to the point where Schwarzschild becomes unstable and new scalarized solutions originate, as well as the behavior of the scalarized black hole branches. We consider first the more conventional case of having $\alpha_2=-1$ and varying $\alpha_3$. The black hole horizon radius and scalar charge as functions of mass are depicted in Fig.~\ref{fig:D_rh_M_var_a3}. The sGB case corresponds to $\alpha_3=0$, and $\alpha_2=-1$ is depicted with a solid dark green line. The rest of the cases with various $\alpha_3$ for which $\alpha_3\dot{\mathcal{G}}_3(\psi_H) \ne \alpha_2 \dot{\mathcal{G}}_2(\psi_H)$ are plotted with solid lines of different colors while a dashed line is used when $\alpha_3\dot{\mathcal{G}}_3(\psi_H) = \alpha_2 \dot{\mathcal{G}}_2(\psi_H)$. As discussed, in the former case the first boundary condition \eqref{eq:BC1} is used, while in the latter case, the second one \eqref{eq:BC2} is employed. Note that because of the choice of $\mathcal{G}_2$ and $\mathcal{G}_3$ in \eqref{eq:G2_G3_choice} the mentioned (in)equality of $\alpha_3\dot{\mathcal{G}}_3$ and $\alpha_2 \dot{\mathcal{G}}_2$ is controlled solely by the values of $\alpha_2$ and $\alpha_3$. Thus, when $\alpha_2\ne \alpha_3$, BC1 is used, while for $\alpha_2 = \alpha_3$ we have to employ BC2. 
	
    In Fig.~\ref{fig:D_rh_M_var_a3} (and other figures below), we plot the Komar mass of the black hole defined as
    \begin{align}\label{eq:KomarMass}
        \mathcal{M} = \frac{1}{2}\lim_{r\to\infty}\left(r^2 \frac{g'_{tt}}{g_{tt}}\right)\,.
    \end{align}
    	
	As one can see, with the increase of $\alpha_3$, the point of bifurcation from the GR branch moves to large masses. We have checked that these points are in agreement with the analytical perturbative results in Eq.~\eqref{eq:BifurcationCondition} for each combination of $\alpha_2$ and $\alpha_3$. Note that this inequality is a sufficient but not necessary condition for instability. Thus, scalarized black holes actually exist also for a little bit larger masses than the mass given by the threshold \eqref{eq:BifurcationCondition}.  
	
	For larger $\alpha_3$ (e.g. $\alpha_3=1.0$ in Fig. \ref{fig:D_rh_M_var_a3} that actually corresponds to the pure Teleparallel case) the branch of scalarized solutions disappears at smaller masses. The end of this sequence resembles a spiraling sequence of branches. Due to numerical difficulties, though, we could build only a small part of them. Such a feature is not unique for the pure Teleparallel case with $\alpha_2=-1.0$ and $\alpha_3=1.0$ and is present also for other large values of $\alpha_3$. Similar spiraling sequences of black hole solutions can be present also for other static non-linearly scalarized black holes in sGB gravity \cite{Blazquez-Salcedo:2022omw}. 
	
	For somewhat smaller $\alpha_3$ (see, for example, $\alpha_3=0.5$ on the figure), the branch already starts behaving a bit differently and it is terminated at a point where the regularity condition related to Eq.~\eqref{eq:BC1} starts being violated. As we decrease $\alpha_3$ further and it gets negative, we observe that the mass of the sequences hits zero at a finite horizon radius. This happens for both BC1 ($\alpha_3=-0.5$ in the figure) and BC2 ($\alpha_3=-1$ in the figure). This is a very peculiar and problematic case. It is well known, though, that Gauss-Bonnet gravity, as well as other quadratic theories of gravity, often poses challenges at small masses, such as loss of hyperbolicity \cite{Blazquez-Salcedo:2018jnn, Ripley:2019hxt,R:2022hlf}, so they should be treated with caution there.
    
    There are even some solutions we found for which a horizon exists for vanishing mass (see the $\alpha_3=-1$ and $\alpha_3 = -0.5$ cases). Let us comment further on this peculiar behavior. In the figures, we plot the Komar mass of the black hole defined by Eq.~eqref{eq:KomarMass} that is related to the asymptotic of the metric function $g_{tt}$ at infinity. Thus, having zero mass simply means that the sign in front of the $1/r$ asymptotic of $g_{tt}$ at infinity switches. This is a non-standard behavior related to the fact that the energy conditions of effective energy-momentum tensor can be violated  locally. This eventually leads to a negative mass. As a matter of fact, solutions with decreasing scalar charges exist also beyond the $M = 0$ point and from a numerical point of view they behave perfectly well. However, such solutions are rather exotic and most probably they are unstable -- that is why we shall not consider them.  Interestingly, such a behaviour, i.e., reaching zero mass at a nonzero horizon radius, has been observed in scalar-curvature theories. The reason for that was the dependence of the ADM mass on the coupling constants of the theory. Such a horizon has been called ``theory horizon'' \cite{Fernandes:2021dsb,Lu:2020iav,Hennigar:2020lsl,Bakopoulos:2022csr}.
    
    Interestingly, despite the fact that the dashed line solutions with $\alpha_3=-1$ employ a rather different boundary condition \eqref{eq:BC2}, they behave qualitatively similar to the boundary condition \eqref{eq:BC1} solutions, and there is a smooth transition between both. In the former case, for a fixed mass, the black holes tend to have a radius of the horizon generally larger compared to Schwarzschild, contrary to the latter cases where the radius of the horizon decreases with respect to GR.

	\begin{figure}[h]
		\includegraphics[scale=0.32]{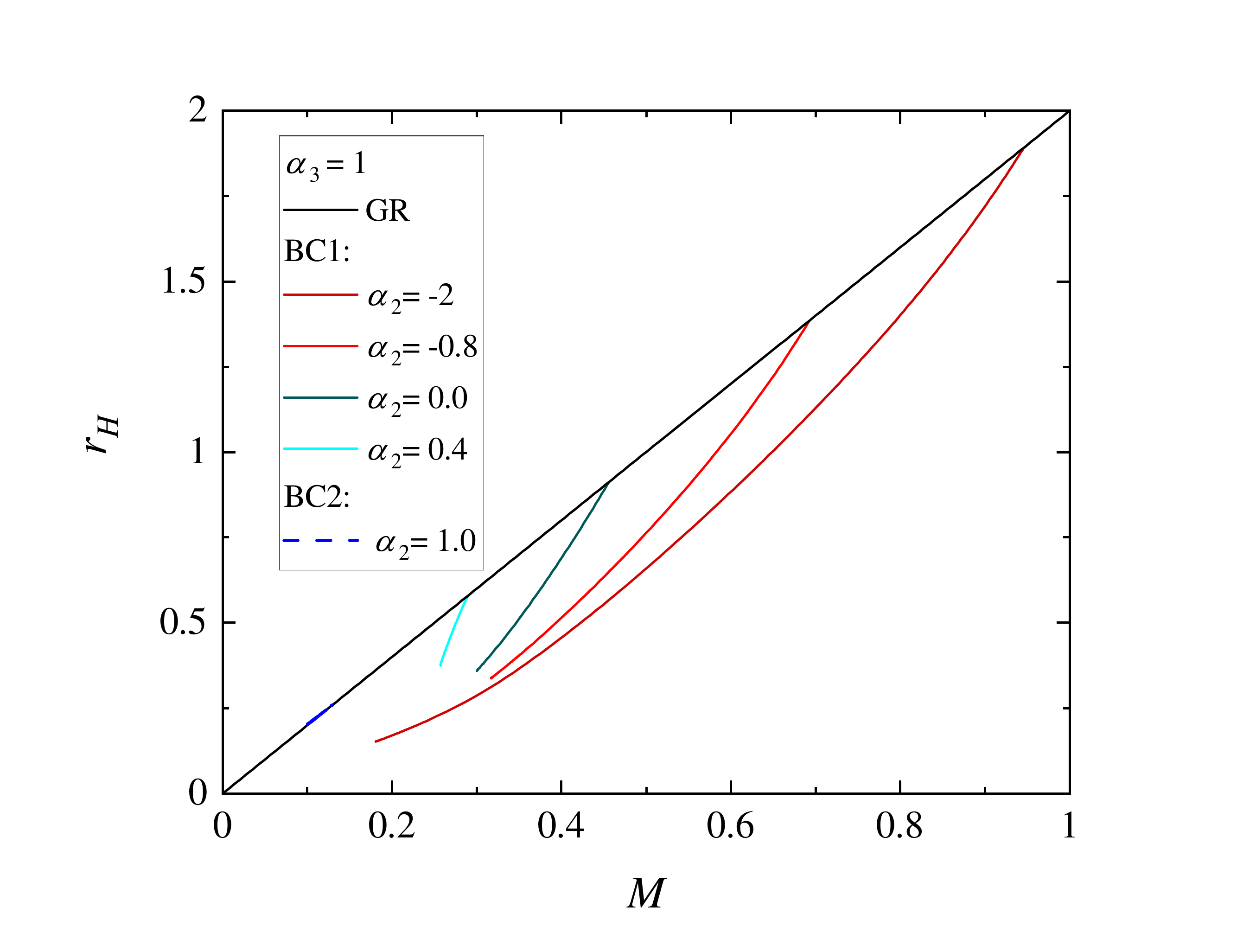}
		\includegraphics[scale=0.32]{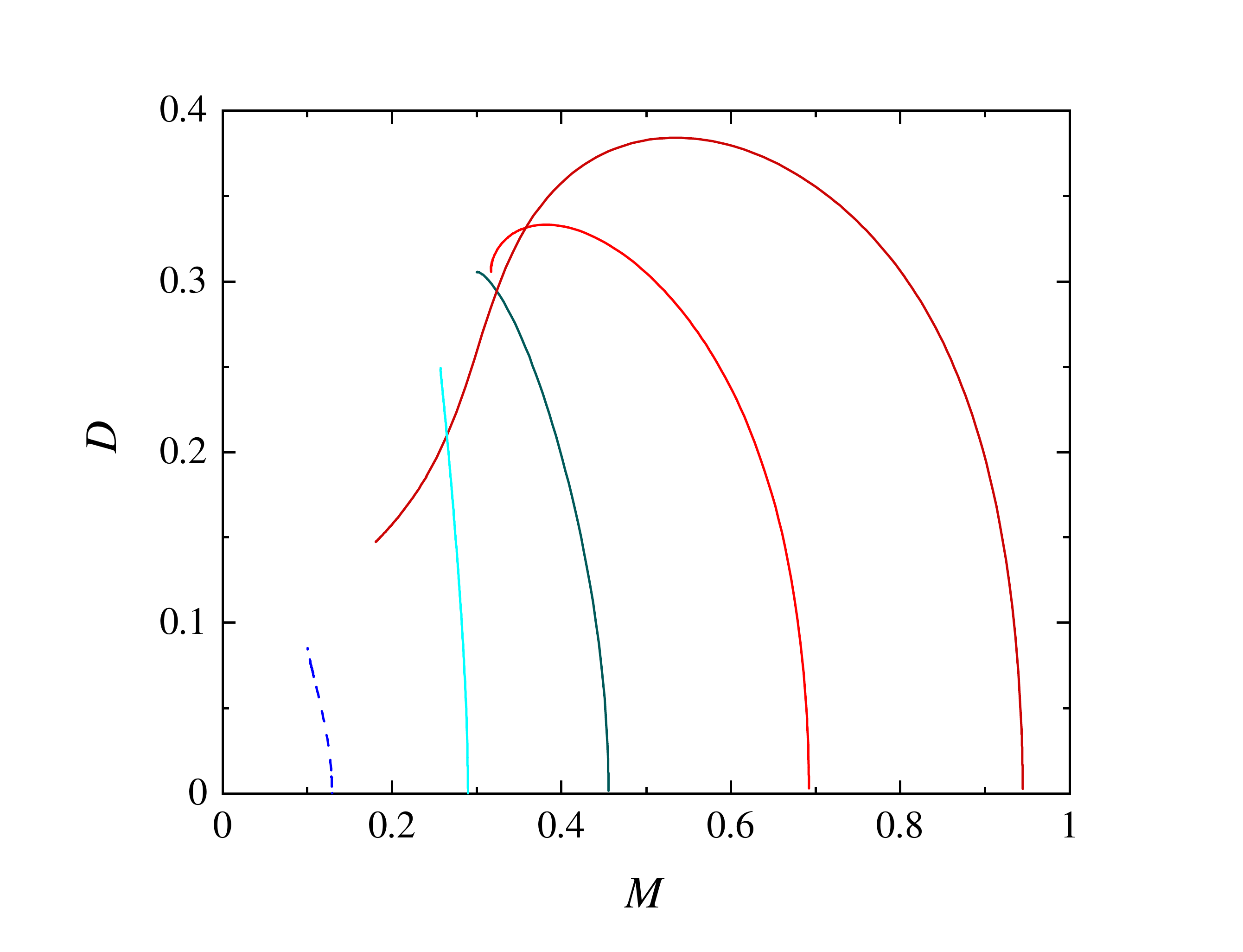}
		\caption{The radius of the horizon (left panel) and the scalar charge (right panel) as functions of mass for a fixed $\alpha_3=1$ and varying $\alpha_2$. The conventions for solution branches are the same as in Fig. \ref{fig:D_rh_M_var_a3}.}
		\label{fig:D_rh_M_var_a2}
	\end{figure}
	
	The case of fixed $\alpha_3=1$ and varying $\alpha_2$ is presented in Fig.~\ref{fig:D_rh_M_var_a2}. Here, contrary to the previous figure, larger $\alpha_2$ move the bifurcation point to smaller masses, something that can be concluded also from Eq.~\eqref{eq:BifurcationCondition}. When $\alpha_2=0$ we have the so-called pure Teleparallel sGB when the Riemannian Gauss-Bonnet term contribution to the field equations is completely turned off. Interestingly, even though this case offers a completely new type of scalarization, the behaviour of the solutions branches is qualitatively very similar to the sGB theory. 
	
	For the considered combinations of parameters, the branches are relatively short, especially for larger positive $\alpha_2$. This is true also for the dashed line branch where the regularity condition at the horizon can always be satisfied (BC2 is employed there). As we comment in the next subsection, the disappearance of solutions at small masses has a different origin for these branches, and the tetrad functions start developing a non-smooth first derivative. We could not give a definite answer, though, whether this is a purely numerical issue or rather an actual peculiarity of the system of equations that prevents the existence of scalarized black holes for smaller masses. 
	
	Note that in the results presented above we were very restrictive in the choice of $\mathcal{G}_2$ and $\mathcal{G}_3$. The observed peculiarities, and especially the disappearance of solutions for small black hole masses, can be controlled by choosing different coupling functions. As it is well known in the sGB gravity, even though the point of bifurcation is a universal property of a system depending only on the leading order expansion of the coupling, the properties and the existence region of the actual nonlinear scalarized black hole solutions can differ drastically for different coupling functions \cite{Doneva:2018rou,Minamitsuji:2018xde,Silva:2018qhn,Doneva:2021tvn}. We have restrained ourselves from studying such a dependence because the main goal of the present paper is to show the existence of scalarized black hole triggered by a Teleparallel term. A detailed study of the possible couplings will be performed elsewhere.
    
    Before going further, let us emphasize again that we focused the discussion on cases where $\alpha_2 = 1$ for various $\alpha_3$ and cases where $\alpha_3=1$ for various $\alpha_2$ since the most important parameter in the discussion of the system is the relative strength between  $\alpha_2$ and $\alpha_3$. The above discussion should thus be representative of the all spectrum of solutions.

	\subsubsection{Radial dependence of the metric and the scalar field}
	After discussing the domain of existence of scalarized solutions, let us turn now to exploring the radial profiles of the tetrad functions and the scalar field. In order to avoid overcrowding with figures we have chosen only some representative combinations of $\alpha_2$ and $\alpha_3$.

\begin{figure}[h]
	\includegraphics[scale=0.32]{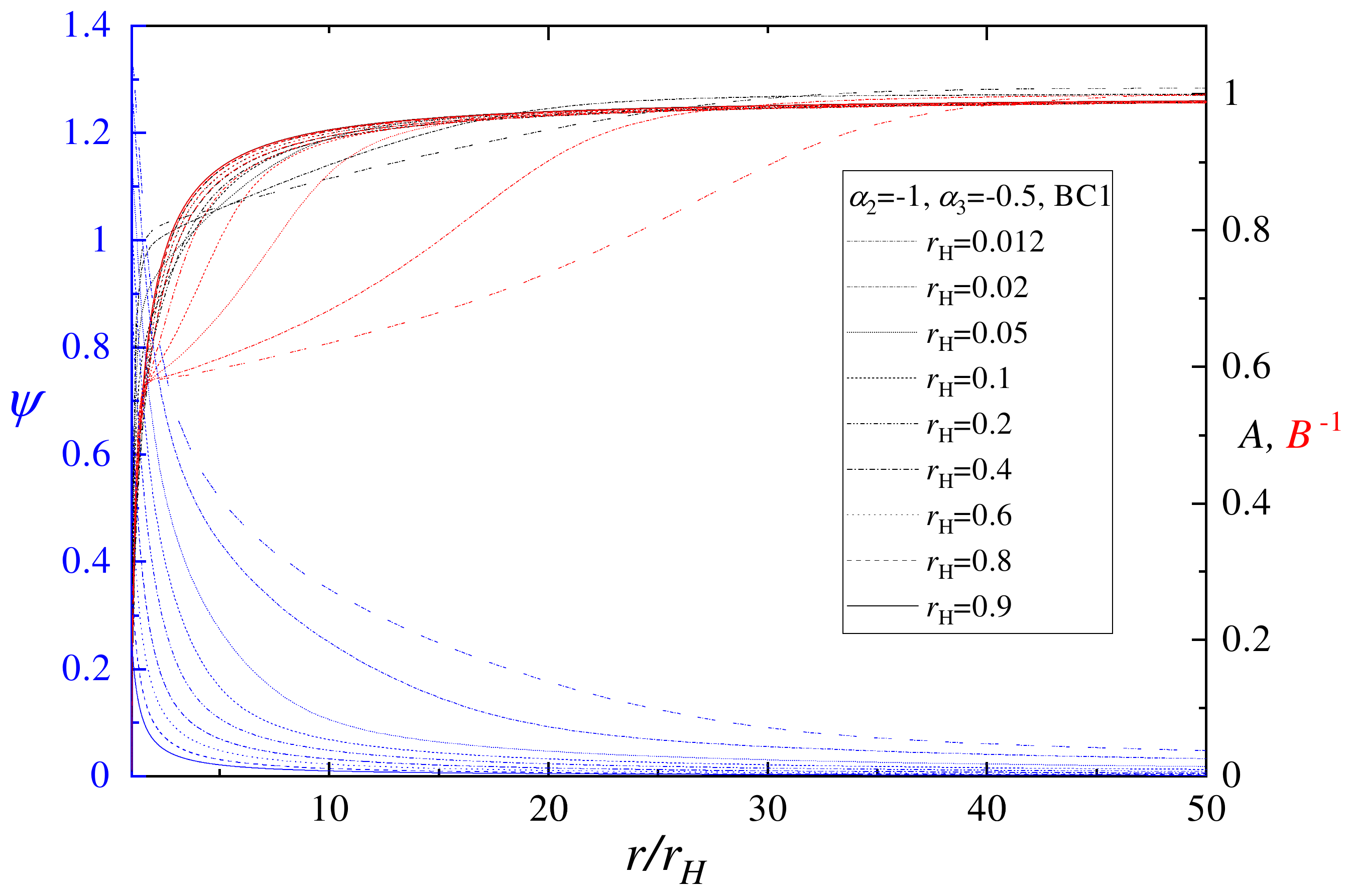}
	\includegraphics[scale=0.32]{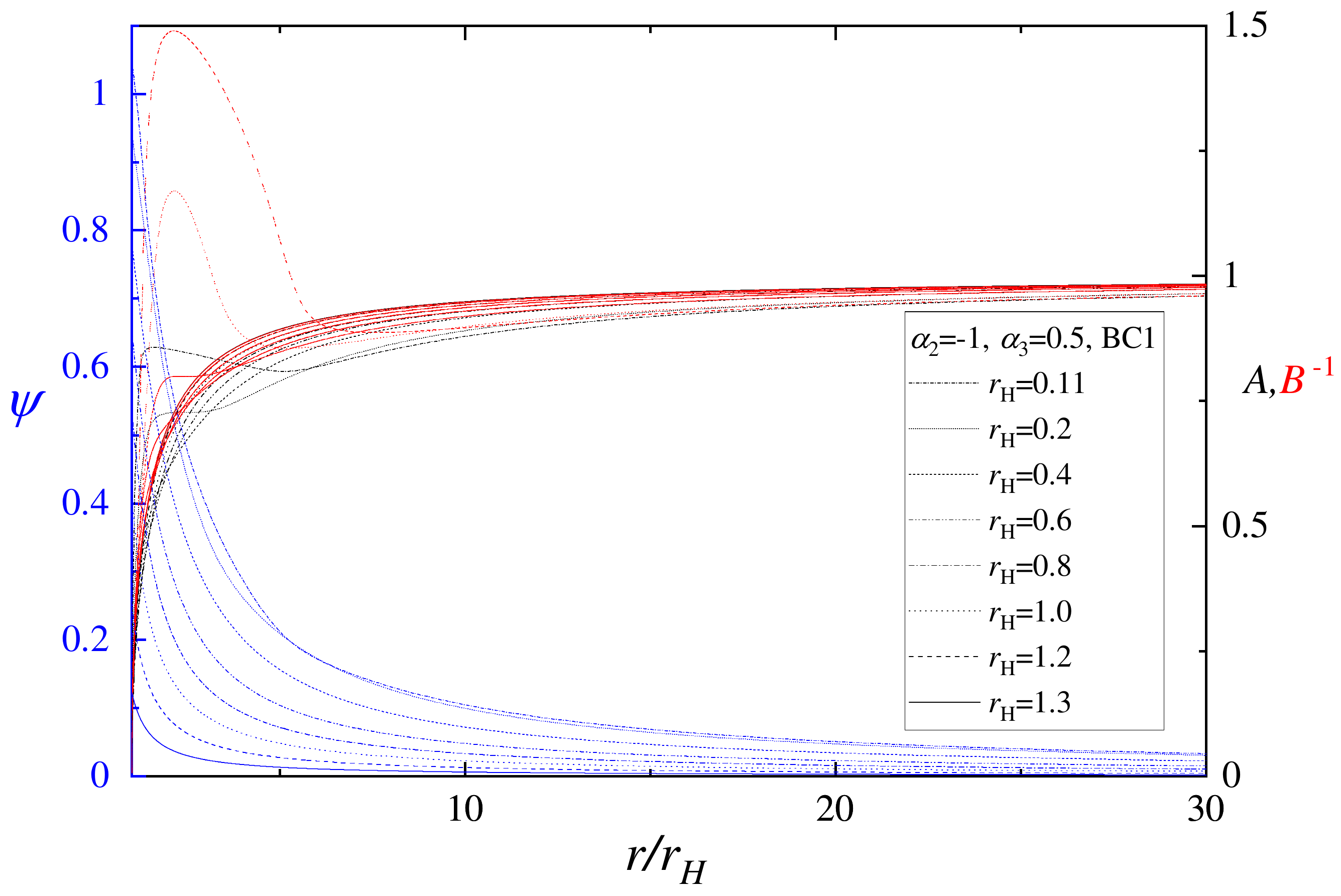}
	\caption{The tetrad functions $A(r)$ and $B^{-1}(r)$, and the scalar field $\psi$ as functions of the normalized to the horizon radius radial coordinate $r/r_H$. The lines corresponding to different functions are marked with different colours -- black, red, and blue for $A(r)$, $B^{-1}(r)$, and $\psi(r)$, respectively. The left $y$-axis corresponds to the $\psi$ function, while the right $y$-axis sets the scale for $A(r)$ and $B^{-1}(r)$. Different patterns of the lines correspond to solutions with varying $r_H$. The theory parameters are fixed to $\alpha_2=-1, a_3=-0.5$ (left panel) and $\alpha_2=-1, a_3=0.5$ (right panel). Since $\alpha_2 \neq a_3$ in both cases, BC1 is used. These solutions are a subset of the ones presented in Fig. \ref{fig:D_rh_M_var_a3}.}
	\label{fig:solutions_BC1}
\end{figure}

	Let us first start with the more conventional set of solutions when $\alpha_2=-1$ and $\alpha_3=\pm 0.5$ and the first boundary condition \eqref{eq:BC1} is employed. The radial profiles of the tetrad functions $A(r)$ and $B^{-1}(r)$, as well as the scalar field $\psi$, for a number of black holes with varying $r_H$ are depicted in Fig.~\ref{fig:solutions_BC1}. As already commented, the tetrad functions are directly related to the metric via $\{g_{tt}(r),-g_{rr}(r)\}=\{A(r)^2,B(r)^2\}$. The selected solutions are a subset of the ones presented in Fig.~\ref{fig:D_rh_M_var_a3} and they are chosen in such a way that $r_H$  roughly spans the range from the bifurcation point up to the leftmost point of the corresponding branch of solutions. These two cases can be considered as a deformation of the sGB scalarized black holes with varying contributions from the Teleparallel gravity.

	For large black holes, both the scalar field and the tetrad functions behave monotonically with the increase of $r$. With the decrease of the horizon radius, though, interesting behavior is observed. Namely, the profile of the tetrad function $A(r)$ starts being deformed developing a rapid change of its derivative close to the horizon and forming a plateau region afterward. For even smaller $r_H$ (see, e.g., $r_H=0.012$ in the right panel of Fig.~\ref{fig:solutions_BC1}), a local maximum of $A(r)$ can form. The $B^{-1}$ function, on the other hand, can develop either a sharp change close to the black hole horizon for small $r_H$ (see, e.g., $r_H=0.11$ in the left panel of Fig.~\ref{fig:solutions_BC1}), or it starts behaving non-monotonically similar to $A$ (the right panel of Fig.~\ref{fig:solutions_BC1}). As a matter of fact, this almost discontinuous behavior of $B'$ for $\alpha_2=-1$ and $\alpha_3=-0.5$ is the reason why we could not calculate solutions with even smaller $r_H$. We could not determine whether such solutions indeed do not exist or if this is a purely numerical problem.  We should note that peculiarities in the metric function behavior can be observed also for small black holes in pure sGB theory \cite{Collodel2022}, even though they are somewhat milder there, e.g. extrema of $A$ do not form for static solutions in the sGB case.   

	Another important feature one can notice in Fig. \ref{fig:solutions_BC1} is that despite the nontrivial behavior for small $r$, both $A(r)$ and $B^{-1}(r)$ tend to be nearly equal as $r$ increases. This is a manifestation of the fact that for large distances we expect from theoretical consideration that the multiplication $A(r)B(r)|_{r\rightarrow\infty} \rightarrow 1$. That has been explicitly checked to be true also for the numerically calculated solutions. Thus, the determinant of the metric is non-zero and non-divergent in this limit.
 
  On the other hand, even though both $A(r)$ and $B^{-1}(r)$ tend to $0$ at the event horizon, this horizon can only be a regular one if $A(r)B(r)|_{r\rightarrow r_H} \rightarrow c \neq 0$ (ensuring again that the metric determinant is neither singular nor vanishing in this limit). We also checked that this was indeed the case for the solutions we obtained numerically, proving that we really have constructed regular and asymptotically flat scalarized black holes.
    
\begin{figure}[H]
		\includegraphics[scale=0.32]{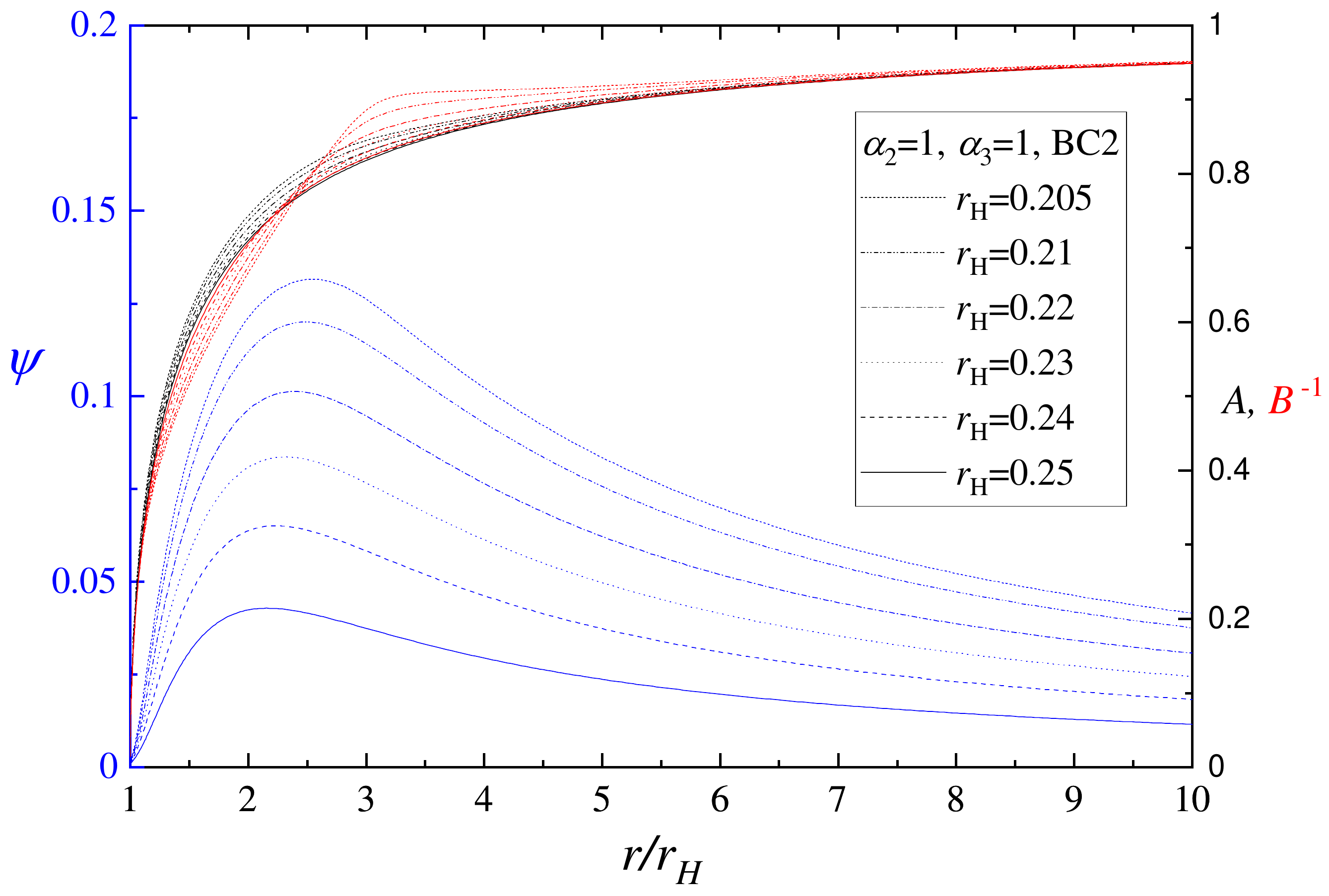}
		\includegraphics[scale=0.32]{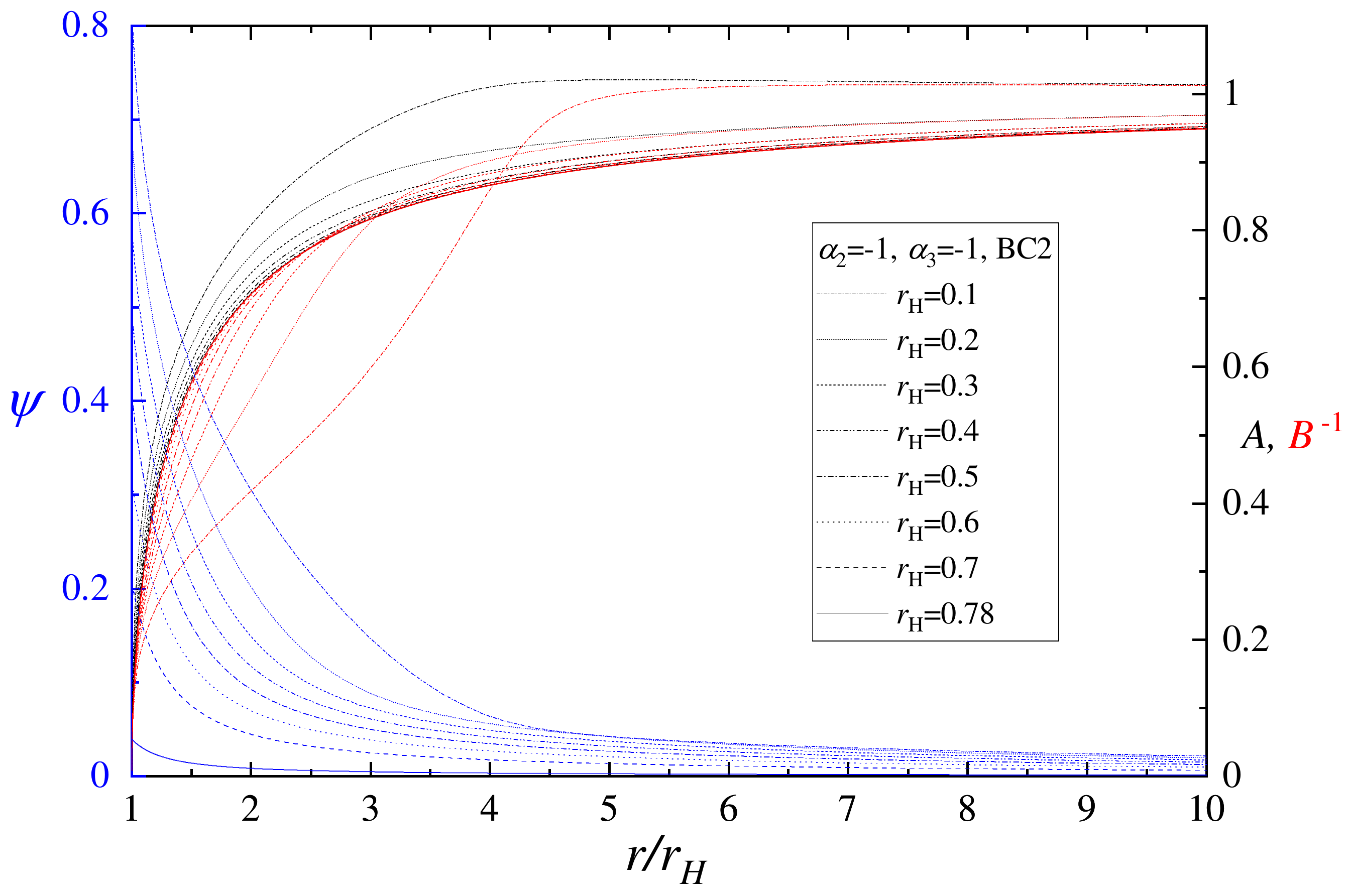}
		\caption{The tetrad functions $A(r)$ and $B^{-1}(r)$, and the scalar field $\psi$ as functions of the normalized to the horizon radius radial coordinate $r/r_H$. Notations are the same as for Fig. \ref{fig:solutions_BC1}, while the theory parameters are set to $\alpha_2=\alpha_3=1$ (left panel) and $\alpha_2=\alpha_3=-1$ (right panel). Since $\alpha_2=\alpha_3$ in both cases, BC2 is used. These solutions are a subset of the ones presented in Figs. \ref{fig:D_rh_M_var_a3} and \ref{fig:D_rh_M_var_a2}.}
		\label{fig:solutions_BC2}
	\end{figure}

    Let us now proceed to solutions with the second boundary condition \eqref{eq:BC2} having  $\alpha_2=\alpha_3$. Up to a normalization constant, the only two possible cases are $\alpha_2=\alpha_3=-1$ and $\alpha_2=\alpha_3=1$ presented in Fig.~\ref{fig:solutions_BC2}. The black hole solutions are a subset of the ones presented in Figs.~\ref{fig:D_rh_M_var_a3} and \ref{fig:D_rh_M_var_a2}, respectively, with $r_H$ spanning the whole range of existence of the corresponding branches.

    Let us first focus on the right panel of Fig.~\ref{fig:solutions_BC2} that resembles more the conventional sGB scalarization where static black holes scalarize only for negative $\alpha_2$. Again, for large $r_H$, the functions $\psi(r)$, $A(r)$ and $B^{-1}(r)$ behave monotonically with $r$. For small black holes, though, it can happen that $A(r)$ and $B^{-1}(r)$ develop  maxima for intermediate $r$ reaching  values larger than one. Afterward, they tend monotonically to $A_\infty=B_\infty=1$. Such behavior leads to a negative mass calculated on the basis of the asymptotic expansions at infinity of $A(r)$ or $B^{-1}(r)$. Remember also that in Fig. \ref{fig:D_rh_M_var_a3} the branches were hitting zero mass at a nonzero $r_H$. This is a manifestation of the same problem. In Fig.~\ref{fig:D_rh_M_var_a3} we have disregarded such negative mass solutions as nonphysical and they are not plotted. 
    
    Now let us proceed with the black hole solutions having $\alpha_2=\alpha_3=1$ and BC2, which are plotted in the left panel of Fig. \ref{fig:solutions_BC2}. Since positive $\alpha_2$ do not allow for scalarization of Schwarzschild black holes in sGB gravity this is another very distinct case\footnote{Note, that scalarization with positive $\alpha_2$ in  sGB gravity is possible but only for rotating black holes, the so-called spin-induced scalarization \cite{Dima:2020yac,Doneva:2020nbb,Herdeiro:2020wei,Berti:2020kgk}.}. Here, the tetrad functions $A(r)$ and  $B^{-1}(r)$ behave monotonically for all solutions. Interestingly, the scalar field is almost vanishing at the black hole horizon, and after reaching a maximum, it starts to decrease monotonically. 
    This is contrary to all other presented cases where the scalar field reaches its maximum value at the horizon. As a matter of fact, such scalar field maxima away from the horizon can be present in sGB gravity but only for rotating solutions up to our knowledge \cite{Doneva:2022yqu}.  
    
\begin{figure}[h!]
		\includegraphics[scale=0.32]{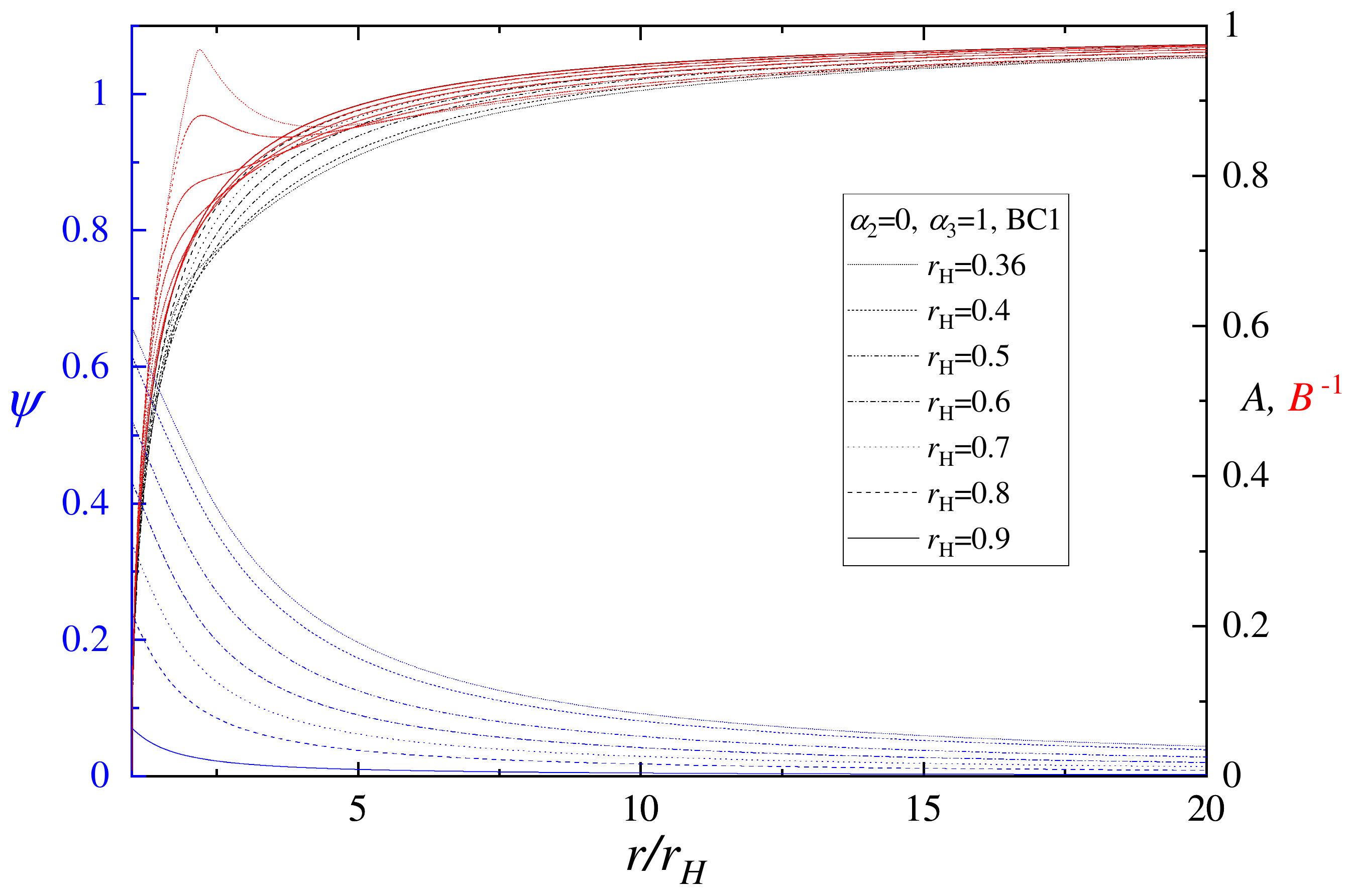}
		\includegraphics[scale=0.32]{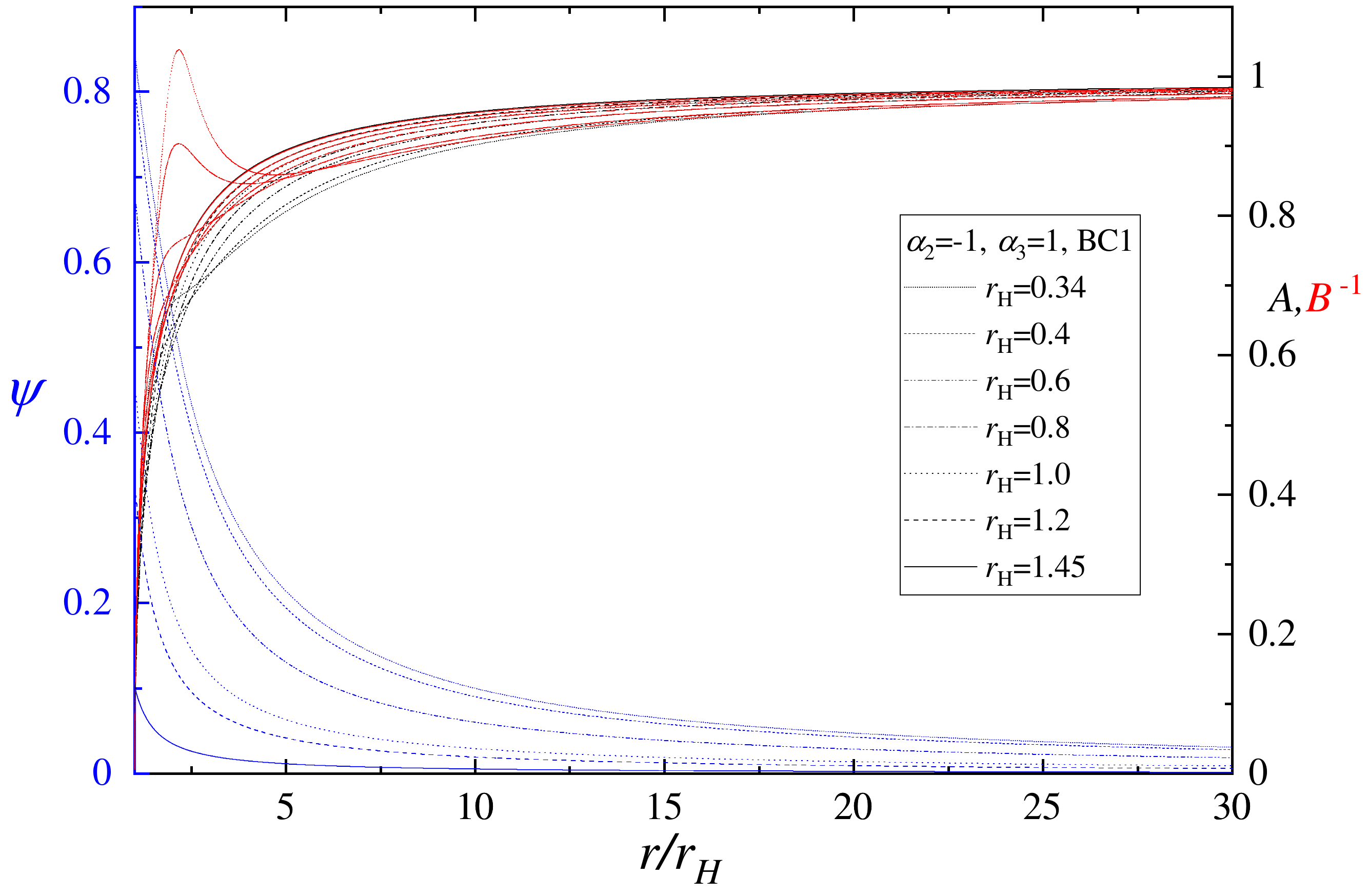}
        \caption{The tetrad functions $A(r)$ and $B^{-1}(r)$, and the scalar field $\psi$ as functions of the normalized to the horizon radius radial coordinate $r/r_H$. Notations are the same as for Fig. \ref{fig:solutions_BC1}. The two figures correspond to the two pure Teleparallel cases with $\alpha_2=0$ and $\alpha_3=1$ (left panel),  and $\alpha_2=-1$ and $\alpha_3=1$ (right panel).  These solutions are a subset of the  ones presented in Figs. \ref{fig:D_rh_M_var_a2} and \ref{fig:D_rh_M_var_a3}, respectively.}
		\label{fig:solutions_BC1_PureTeleparallel}
	\end{figure}

	The last cases we present are the pure Teleparallel scalarization with either $\alpha_2=0$ and $\alpha_3=1$, or $\alpha_2=-1$ and $\alpha_3=1$, as discussed in the previous section. Selected representative solutions are plotted in Fig. \ref{fig:solutions_BC1_PureTeleparallel}. Despite the fact that here the Riemannian Gauss-Bonnet term is missing and only the Teleparallel one contributes, the qualitative behavior is similar to the rest of the solutions. For sufficiently large $r_h$, both the scalar field and the tetrad functions $A(r)$ and $B^{-1}(r)$ are monotonic. $A(r)$ starts being more deformed for small $r_H$ where the branch of scalarized solutions is terminated, while $B^{-1}(r)$ can even develop maxima close to the horizon for small mass black holes. 
	
\begin{figure}[h!]
        \includegraphics[scale=0.25]{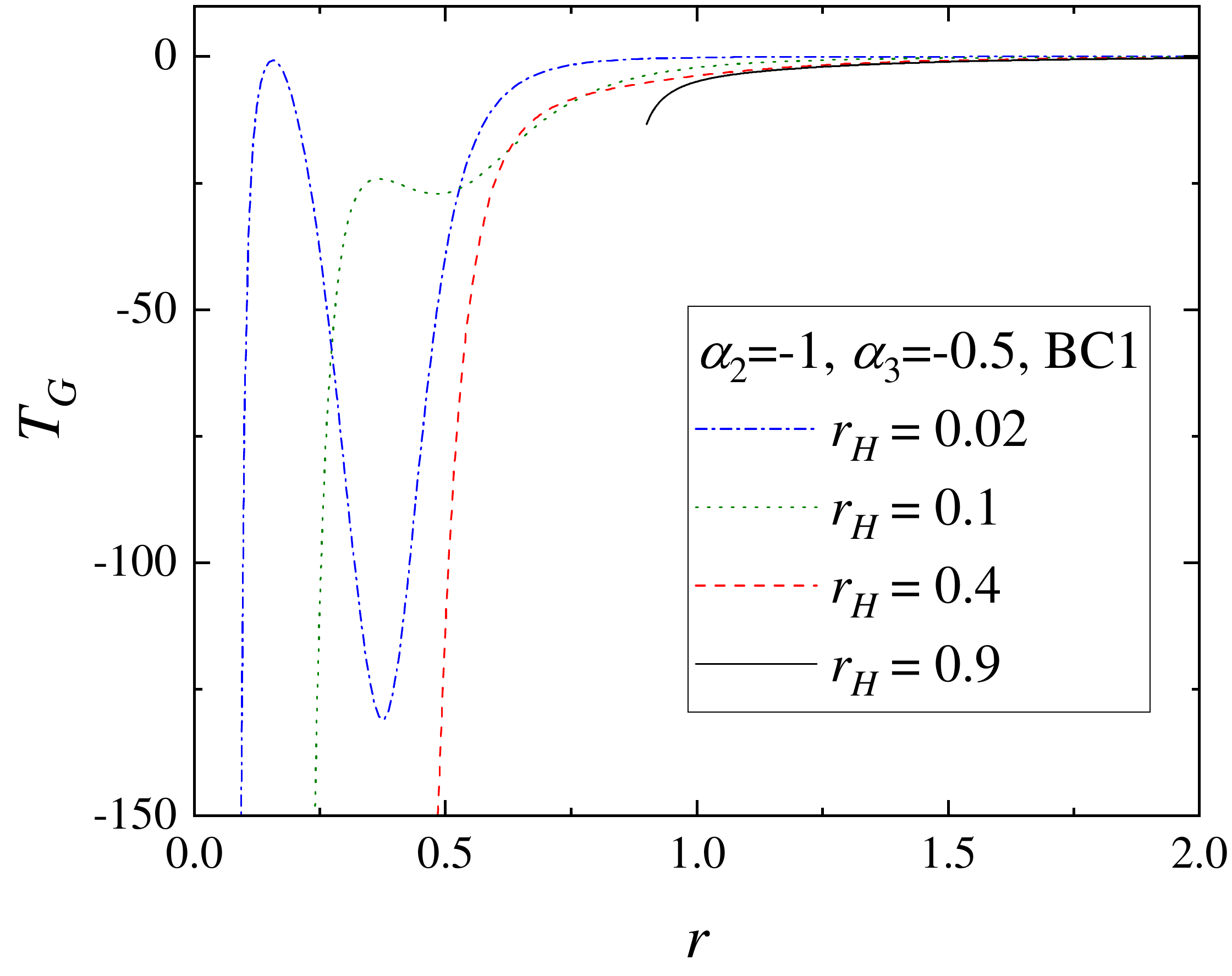}
        \includegraphics[scale=0.25]{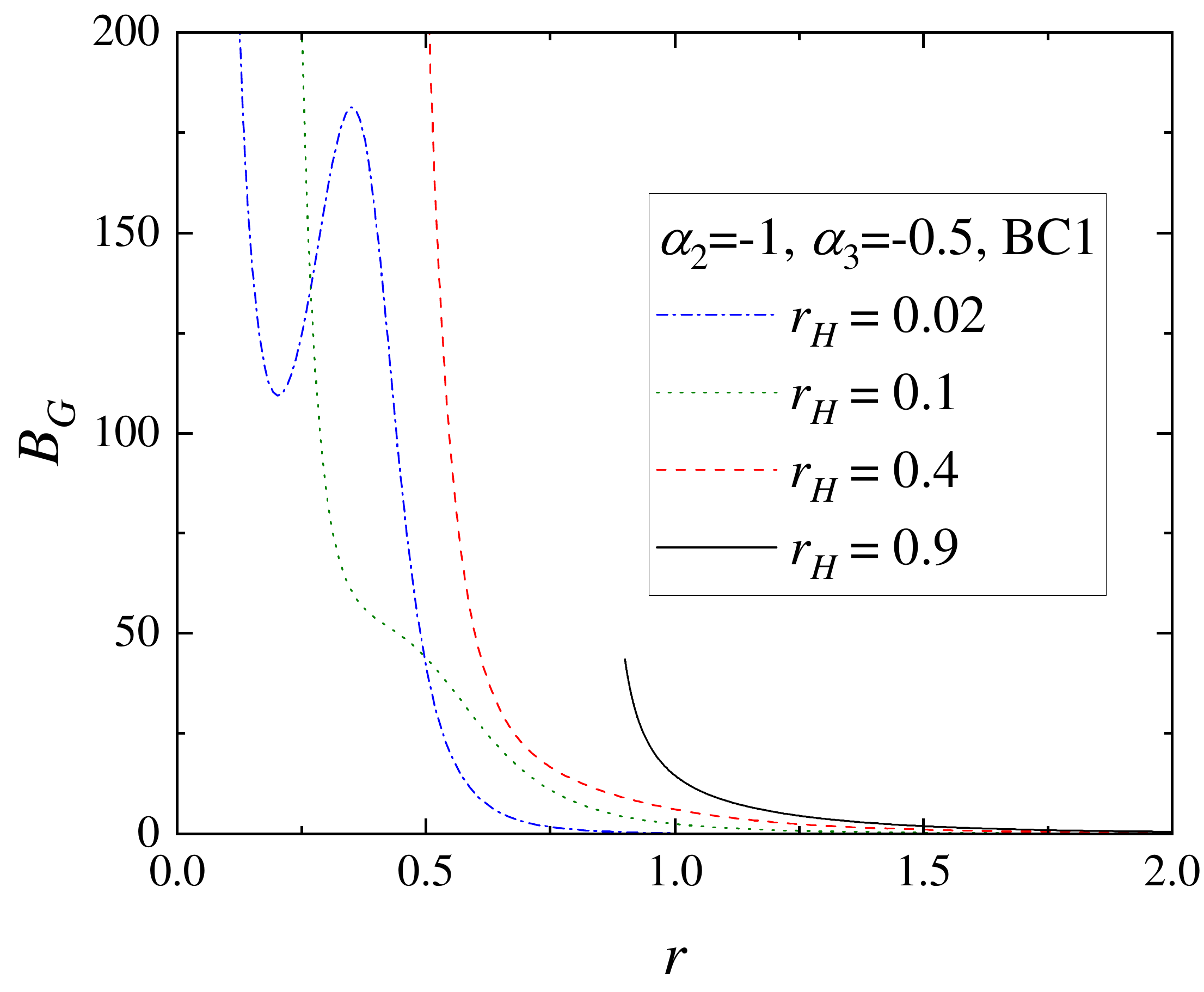}
        \includegraphics[scale=0.25]{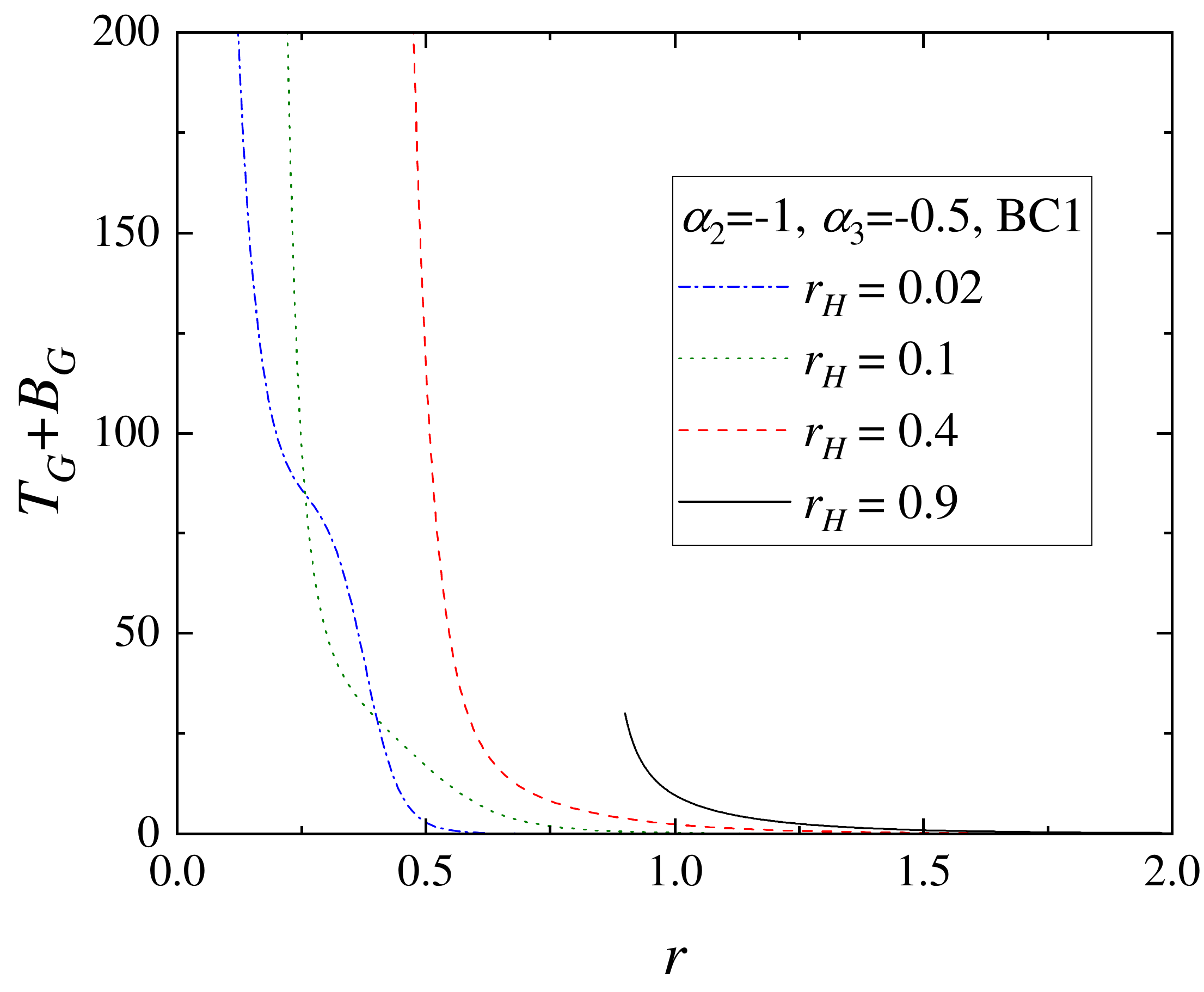}
        \caption{The teleparallel terms $T_G(r)$ and  $B_G(r)$, as well as the teleparallel Gauss-Bonnet invariant equal to the summation of both (see Eq.~\eqref{GB})  as functions of radial coordinate $r$. The coupling parameters are chosen to be $\alpha_2=-1$ and $\alpha_3=-0.5$. These solutions are a subset of the ones presented in Fig. \ref{fig:D_rh_M_var_a3}.}
        \label{fig:solutions_TGBG}
    \end{figure}

    It would be interesting to examine as well the behavior of the different teleparallel terms $T_G$ and $B_G$. In Fig. \ref{fig:solutions_TGBG} we plot these quantities together with the teleparallel Gauss-Bonnet invariant $\lc{G}= T_{G}+B_G$ for a representative combination of parameters $\alpha_2=-1$ and $\alpha_3=-0.5$ (the sequence of solutions can be found in Fig. \ref{fig:D_rh_M_var_a3}). As seen, for larger $r_H$ the teleparallel quantities behave monotonically with the increase of $r$. For small $r_H$ black holes, though, local extrema can form that is an indirect consequence of the on-monotonous behavior of the metric functions presented above. 

    We like to end this section by highlighting our finding that already a non-minimal coupling to one of the purely teleparallel terms $T_G$ or $B_G$, which compose the Riemannian Gauss-Bonnet term $\lc{G}$, suffices to obtain spontaneous scalarization. The individual terms $T_G$ and $B_G$ can only be expressed in terms of torsion and not in terms of a metric or curvature. Thus, we demonstrated that spontaneous scalarization can be sourced by torsion, which in the Riemannian case, is usually set to zero, and not necessarily by the curvature of spacetime.   
    An immediate, very interesting question, that arises is, if the origin of spontaneous scalarization lies in the properties of non-Riemannian aspects (here torsion) of the spacetime geometry. The investigation of this question paves the way to a better understanding of the physical consequences of deviations from a purely Riemannian spacetime geometry.

\section{Summary and Conclusion}\label{sec:concl}
The sGB theory has attracted a lot of attention due to the fact that it can naturally evade the no-scalar-hair theorems, and it is motivated by the attempts to quantize gravity. A boost in the field in the last years came from the discovery that in the weak field regime, it can describe the Schwarzschild solution while in the strong field regime, one can obtain deviations from it in a mechanism labeled spontaneous scalarization. Usually, this process has been studied in sGB gravity and in other Riemannian scalar-tensor theories but up to our knowledge, there are no such studies concerning non-Riemannian theories of gravity, involving torsion or non-metricity. Let us quickly summarize our findings and conclude.

\subsection{Summary}
In this paper, we have considered a Teleparallel theory where the general curvature is vanishing, the metric compatibility condition holds and the torsion tensor is non-vanishing and encodes the gravitational dynamics. 
Since one can formulate an equivalent theory to GR in the Teleparallel framework, which is TEGR, one can then introduce a scalar field and follow a similar approach as in the Riemannian case, which is to construct theories with non-minimal couplings between a scalar field and the gravitational sector. It turns out that there exists an analogue Teleparallel Gauss-Bonnet invariant that contains two parts: The first one is a topological invariant in 4-dimensions ($T_G$ defined in~\eqref{TG}) and the second one is a boundary term in any dimension ($B_G$ defined in~\eqref{BG}). Those Teleparallel invariants can be directly linked to the Riemannian Gauss-Bonnet invariant through the relationship~\eqref{GB}. We then construct a Teleparallel sGB theory constructed by couplings between the scalar field and $T_G$ and $B_G$ that (due to~\eqref{GB}) can be rewritten as~\eqref{action2F} containing a contribution from the sGB action plus an additional Teleparallel term, which would give a richer phenomenology. 

In the Teleparallel framework, one can construct fully invariant theories, i.e., invariant under both diffeomorphism and local Lorentz transformations~\cite{Krssak:2018ywd,Bahamonde:2021gfp}. This is done by considering the flat spin connection and the tetrad fields as dynamical variables. However, one can always choose a gauge such that the spin connection is vanishing, which is the gauge that we used during this paper for studying spherical symmetry. By imposing the condition that the torsion respects the same symmetries as the metric, the corresponding form of the tetrad could have a general form expressed in~\eqref{sphtetrad}. By substituting this tetrad in the antisymmetric equations of our theory, we found two possible generic tetrad solutions giving the same form of the metric. This means that our theory would have two sets of spherically symmetric field equations, depending on the branch chosen. We concentrated on the second one where the tetrad is a complex one~\eqref{tetrad2}.

The most important part of our paper was presented in Sec.~\ref{sec:5} where we studied the spherically symmetric complex tetrad equations for Teleparallel sGB gravity and find scalarized black hole solutions. We first concentrated on the analytical part of the equations and find perturbed solutions around Schwarzschild in Sec.~\ref{sec:perturbed} which contains as a special case the sGB perturbed solution found previously in~\cite{Bryant:2021xdh}. We then studied the behaviour of the equations near the horizon and at infinity in Sec.~\ref{sec:BC}, finding that there are two different branches of boundary conditions ensuring the regularity of the scalar field's first derivative at the event horizon depending on the theory chosen. The first one given by~\eqref{eq:BC1} is a generalization of the sGB boundary condition for the scalar field at the horizon while the second one~\eqref{eq:BC2} is a new branch that only exists in the Teleparallel framework and does not contain any square root. We finalized the analytical study by studying the spontaneous scalarization mechanism in Sec.~\ref{sec:spo} where we found the bifurcation bound~\eqref{eq:BifurcationCondition} which provides a sufficient condition to have unstable modes for the Schwarzschild background. That bound again generalizes the well-known sGB result obtained in~\cite{Doneva:2017bvd}. 

In Sec.~\ref{sec:num} we explicitly constructed asymptotically flat scalarized black hole solutions using numerical techniques for solving the full nonlinear set of equations. Due to the richness of the theory, we have focused mainly on proving the existence of such solutions and exploring their basic properties in the case of a coupling of the form \eqref{eq:G2_G3_choice} rather than a detailed exploration of the parameters space. We examined all possible interesting limiting cases, which include the pure Teleparallel cases without contribution from the pure Riemannian Gauss-Bonnet term, as well as the solutions with both boundary conditions discussed in the previous paragraph. We found that in a certain parameter range when the contribution of the pure Teleparallel term is sub-dominant, the  behaviour of the solutions is qualitatively the same as in the sGB case. This might change, though, when the pure Teleparallel term starts having a significant contribution.

Interestingly, the transition between different regimes and sectors of the theory happens smoothly, even in cases when one has to switch between the two different boundary conditions. We could observe very interesting non-monotonic behavior of the metric functions close to the horizon for small mass black holes that form local minima and maxima. This is something completely not typical for static black hole solutions and can have very interesting implications, especially for orbits around such objects. The small mass region is often problematic, though, in Gauss-Bonnet type of theories and loss of hyperbolicity is often observed there \cite{Blazquez-Salcedo:2018jnn, Ripley:2019hxt,R:2022hlf}, at least in the sGB case. That is why, in order to answer the question of whether such effects can have astrophysical relevance, one should study the black hole dynamics. This is a study underway. 

Another interesting fact is that, in the considered theory, it is possible to have static black hole scalarization for coupling functions for which static black holes in sGB gravity can not scalarize but instead spin-induced scalarization is present. This is a very interesting observation because it turns out that, in this case, it is possible to have non-monotonic behavior of the scalar field close to the horizon even for the fundamental nodeless scalar field branch of the black hole. Until now similar non-monotonicity in sGB gravity was observed only for rotating black holes. 

\subsection{Conclusion}
To conclude, we have derived novel scalarized black hole solutions, sourced by torsion, that have not been found nor reported in the past elsewhere (see the review~\cite{Doneva:2022ewd} for a compilation of the scalarization process in Riemannian theories of gravity). Even though we restricted ourselves to static spherically symmetric solutions, the black holes share a lot of similarities not only with the static sGB case but also with the rotating sGB black holes. On the other hand, all sectors of the Teleparallel gravity we considered are well-behaved numerically. Most interestingly, we found that already coupling one of the two Teleparallel contributions $T_G$ and $B_G$, which form the Riemannian Riemannian Gauss-Bonnet term $\lc{G}$, suffices to trigger scalarization. This observation leads to the question if torsion (or other properties of non-Riemannian geometry) might fundamentally be the origin of the scalarization properties. Important additional studies, in order to understand the properties of the teleparallel scalarized BHs further in the future, are: their stability, their thermodynamic properties (for this in particular, the study of BH thermodynamics in teleparallel gravity has to be further developed \cite{Blagojevic:2019gsd}), their dynamics in the perturbative and the nonlinear regime, and if they suffer from loss of hyperbolicity similar to the sGB case or not.

Our study can open a new unexplored window in the study of scalarized black hole solutions in non-Riemannian theories of gravity. We have just shown the existence of solutions using the complex tetrad, but one could also perform a similar analysis with the real tetrad~\eqref{tetrad1} and study its differences. Further, we considered an exponential type of coupling but one could consider other coupling functions (such as polynomial or lineal couplings), as well as other theory parameters, where solutions exist already in the sGB theory. 

These studies should be extended also to spontaneous scalarization of neutron stars to realize whether our theory could describe other astrophysical scenarios. Furthermore, since our formulated theory contains the sGB theory as a special case, one might expect that axially symmetric solutions would also exist but with a larger range of parameters.

\section*{Acknowledgements}
SB is supported by JSPS Postdoctoral Fellowships for Research in Japan and KAKENHI Grant-in-Aid for Scientific Research No. JP21F21789.  DD acknowledges financial support via an Emmy Noether Research Group funded by the German Research Foundation (DFG) under grant no. DO 1771/1-1. S.Y. acknowledges the support by the National Science Fund of Bulgaria under Contract No. KP-06-H28/7. L.D. is supported by a grant from the Transilvania Fellowship Program for Postdoctoral Research/Young Researchers (September 2022).  CP is funded by the Deutsche Forschungsgemeinschaft (DFG, German Research Foundation) - Project Number 420243324 and acknowledges the excellence cluster QuantumFrontiers funded by the Deutsche Forschungsgemeinschaft (DFG, German Research Foundation) under Germany’s Excellence Strategy – EXC-2123 QuantumFrontiers – 390837967. SB and CP would like to acknowledge networking support by the COST Action CA18108.
	
\bibliographystyle{utphys}
\bibliography{references}
	
	\appendix
	\section{Perturbed solutions}\label{appendix1}
	The functions appearing in the perturbed solutions of Sec.~\ref{sec:perturbed} (Eqs.~\eqref{sola}-\eqref{solc}) are given by
	\begin{eqnarray}
		\tilde{a}_1(r)&=&-\frac{1}{M^6 r^2}-\frac{1}{M^5 r^3}-\frac{52}{3 M^4 r^4}-\frac{2}{M^3 r^5}-\frac{16}{5 M^2 r^6}+\frac{368}{3 M r^7}\,,\\
		\tilde{a}_2(r)&=&\frac{2}{M^6 r^2}+\frac{18}{M^5 r^3}+\frac{8}{3 M^4 r^4}+\frac{4}{M^3 r^5}-\frac{1248}{5 M^2 r^6}+\frac{736}{3 M r^7}\,,\\
		\tilde{a}_3(r)&=&-\frac{1}{M^6 r^2}-\frac{17}{M^5 r^3}+\frac{44}{3 M^4 r^4}+\frac{126}{M^3 r^5}-\frac{1232}{5 M^2 r^6}+\frac{368}{3 M r^7}\,,\\
		\tilde{b}_1(r)&=&-\frac{1}{M^6 r^2}-\frac{1}{M^5 r^3}-\frac{52}{3 M^4 r^4}-\frac{2}{M^3 r^5}-\frac{16}{5 M^2 r^6}+\frac{368}{3 M r^7}\,,\\
		\tilde{b}_2(r)&=&-\frac{1}{M^6 r^2}-\frac{17}{M^5 r^3}+\frac{44}{3 M^4 r^4}+\frac{126}{M^3 r^5}-\frac{1232}{5 M^2 r^6}+\frac{368}{3 M r^7}\,,\\
		\tilde{b}_3(r)&=&\frac{2}{M^6 r^2}+\frac{18}{M^5 r^3}+\frac{8}{3 M^4 r^4}+\frac{4}{M^3 r^5}-\frac{1248}{5 M^2 r^6}+\frac{736}{3 M r^7}\,,\\
		\tilde{\psi}_1(r)&=&\frac{2}{M^3 r}+\frac{2}{M^2 r^2}+\frac{8}{3 M r^3}\,,\\
		\tilde{\psi}_2(r)&=&-\frac{2}{M^3 r}-\frac{2}{M^2 r^2}+\frac{8}{3 M r^3}\,,\\
		\tilde{\psi}_3(r)&=&-\frac{73}{30 M^9 r}-\frac{73}{30 M^8 r^2}-\frac{146}{45 M^7 r^3}-\frac{73}{15 M^6 r^4}-\frac{224}{75 M^5 r^5}-\frac{16}{9 M^4 r^6}\,,\\
		\tilde{\psi}_4(r)&=&-\frac{41}{30 M^9 r}-\frac{41}{30 M^8 r^2}-\frac{82}{45 M^7 r^3}+\frac{13}{5 M^6 r^4}+\frac{64}{25 M^5 r^5}-\frac{16}{9 M^4 r^6}\,,\\
		\tilde{\psi}_5(r)&=&\frac{53}{30 M^9 r}+\frac{53}{30 M^8 r^2}+\frac{106}{45 M^7 r^3}-\frac{9}{5 M^6 r^4}-\frac{32}{25 M^5 r^5}-\frac{16}{9 M^4 r^6}\,,\\
		\tilde{\psi}_6(r)&=&\frac{53}{30 M^9 r}+\frac{53}{30 M^8 r^2}+\frac{106}{45 M^7 r^3}+\frac{53}{15 M^6 r^4}+\frac{64}{75 M^5 r^5}-\frac{16}{9 M^4 r^6}\,.
	\end{eqnarray}
\end{document}